\documentclass[fleqn,usenatbib]{mnras}

\usepackage{newtxtext,newtxmath}
\usepackage[T1]{fontenc}

\DeclareRobustCommand{\VAN}[3]{#2}
\let\VANthebibliography\thebibliography
\def\thebibliography{\DeclareRobustCommand{\VAN}[3]{##3}\VANthebibliography}

\usepackage{graphicx}	
\usepackage{amsmath}	
\usepackage[version=4]{mhchem}
\usepackage{hyperref}
\usepackage[separate-uncertainty=true]{siunitx}
\usepackage{bm}
\usepackage{mismath}
\usepackage{lipsum}  
\usepackage{float}

\renewcommand{\d}[1]{\ensuremath{\operatorname{d}\!{#1}}}
\DeclareSIUnit\bar{bar}
\DeclareSIUnit\AU{AU}
\DeclareSIUnit\dex{dex}
\DeclareSIUnit\erg{erg}
\DeclareSIUnit\day{day}
\DeclareSIUnit\year{yr}
\newcommand{\WPMS}{\watt\per\meter\squared}
\newcommand{\GPa}{\giga\pascal}
\newcommand{\IW}{\Delta\text{IW}}

\title[Tidal heating and the L 98-59 system]{Self-limited tidal heating and prolonged magma oceans in the L 98-59 system}

\author[H. Nicholls et al.]{
Harrison Nicholls,$^{1}$\thanks{E-mail: harrison.nicholls@physics.ox.ac.uk}
Claire Marie Guimond,$^{1}$
Hamish C. F. C. Hay,$^{2}$
Richard D. Chatterjee,$^{1}$
\newauthor
Tim Lichtenberg,$^{3}$
and Raymond T. Pierrehumbert$^{1}$
\\
$^{1}$Atmospheric, Oceanic, and Planetary Physics, Department of Physics, University of Oxford, Oxford OX1 3PU, United Kingdom \\
$^{2}$Department of Earth Sciences, University of Oxford, Oxford OX1 3AN, United Kingdom \\
$^{3}$Kapteyn Astronomical Institute, University of Groningen, P.O. Box 800, 9700 AV Groningen, The Netherlands
}

\date{Accepted XXX. Received YYY; in original form ZZZ}
\pubyear{\the\year{}}
\begin{document}
\label{firstpage}
\pagerange{\pageref{firstpage}--\pageref{lastpage}}
\maketitle

\begin{abstract}
Rocky exoplanets accessible to characterisation often lie on close-in orbits where tidal heating within their interiors is significant, with the L 98-59 planetary system being a prime example.  As a long-term energy source for ongoing mantle melting and outgassing, tidal heating has been considered as a way to replenish lost atmospheres on rocky planets around active M-dwarfs. We simulate the early evolution of L 98-59 b, c and d using a time-evolved interior-atmosphere modelling framework, with a self-consistent implementation of tidal heating and redox-controlled outgassing.  Emerging from our calculations is a novel self-limiting mechanism between radiative cooling, tidal heating, and mantle rheology, which we term the `radiation-tide-rheology feedback'. Our coupled modelling yields self-limiting tidal heating estimates that are up to two orders of magnitude lower than previous calculations, and yet are still large enough to enable the extension of primordial magma oceans to Gyr timescales. Comparisons with a semi-analytic model demonstrate that this negative feedback is a robust mechanism which can probe a given planet's initial conditions, atmospheric composition, and interior structure. The orbit and instellation of the sub-Venus L 98-59 b likely place it in a regime where tidal heating has kept the planet molten up to the present day, even if it were to have lost its atmosphere. For c and d, a long-lived magma ocean can be induced by tides only with additional atmospheric regulation of energy transport.
\end{abstract}



\section{Introduction}
\label{sec:intro}

The interiors of rocky planets are thought to be initially molten due to the large energy released during accretion \citep{abe_early_1986, rubie_formation_2007, warren_moon_1985, lichtenberg_review_2025}. These early `magma oceans' subsequently cool through thermal emission, and in many cases solidify (e.g. Earth and Venus). However, in some cases the interiors of rocky planets could be sustained in a permanently molten state by a continuous input of energy sufficient to balance radiative cooling to space. This energy balance is modulated by the radiative properties of any overlying atmosphere \citep{elkins_linked_2008, hamano_lifetime_2015, nicholls_redox_2024}. A trivial case of a permanent magma ocean is where heating by a planet's host star is sufficient to raise the equilibrium surface temperature of the planet beyond the solidus of its mineral assemblage \citep{meier_interior_2023,boukare_deep_2022}. One such case is the super-Earth 55 Cancri e, for which observations have raised the possibility for the presence of a volatile-rich atmosphere in outgassing equilibrium with a magma ocean \citep{hu_55cnc_2024, demory_55cnc_2016, patel_55cnc_2024}. Another potential factor in a planet's energy budget is the interior dissipation of heat generated by tidal stresses from planet-planet and planet-star gravitational interactions \citep{driscoll_tidal_2015, hay_tides_2019, quick_forecasting_2020}. Sufficiently large tidal heating within the interiors of close-in exoplanets could keep their mantles molten or allow for ongoing volcanic activity \citep{kaltenegger_volcanism_2010, henning_tidally_2009, seligman_potential_2024, ostberg_prospect_2023}. The total power $\dot{E}$ dissipated by tides within a planet is known to strongly scale with a its radius $R_p$, orbital period $P$, and eccentricity $e$: $\dot{E} \propto R_p^5 P^{-5}e^2$ \citep[e.g.,][]{segatz_io_1988}. 
\par 
Famously, heating within the interior of Io is thought to be a key driver of ongoing volcanic activity \citep{peale_io_1979, segatz_io_1988, matsuyama_io_2022}. Time-varying tides within Io are induced by its eccentric orbit around Jupiter, enabled by gravitational interactions with its sister moons Europa and Ganymede. The tides acting on Io are therefore analogous to tides raised within a quasi-synchronously rotating planet in an eccentric orbit around its host star. Radioactivity and electromagnetic induction are additional sources of interior heating \citep{lodders_planetary_1998, Kislyakova_induction_2017, Kislyakova_induction_2023}. However, recent measurements by the Juno and Galileo probes suggest that Io does not maintain a permanent subsurface magma ocean \citep{park_io_2024}.
\par 
L 98-59 is a bright M3V star which has been confirmed to host three exoplanets, with tentative detections of two more \citep{demangeon_warm_2021, kostov_l9859_2019}. Observations of the three confirmed planets (L 98-59 b, c, and d) have been made with \textit{TESS}, \textit{JWST}, \textit{HST}, \textit{HARPS}, and \textit{ESPRESSO}. \textit{HST} transit depth constraints placed on the innermost planet (b, a sub-Venus) rule out an extended envelope \citep{demangeon_warm_2021, zhou_hubble_2022, damiano_transmission_2022}. Recent observations with \textit{JWST} NIRSpec \citep{belloarufe_evidence_2025} have been used to infer an \ce{SO2}-rich secondary atmosphere on planet b, suggested to be supplied by outgassing from an oxidised $f\ce{O2} \gtrsim \IW+2.7$ \textit{subsurface} magma ocean permanently sustained by tidal heating. Although an \ce{SO2}-rich atmosphere is favoured, a `wide range of lower \ce{SO2} abundances' and a bare rock are also consistent with their observations of planet b. The presence of hazes/clouds are disfavoured for this planet. Observational constraints on the second planet (c, a super-Earth) also disfavour \ce{H2}-dominated atmospheres, suggesting that it may host a high molecular weight secondary atmosphere or no atmosphere at all \citep{scarsdale_compass_2024, zhou_hubble_2023, barclay_transmission_2023}. In comparison, transmission spectroscopy of planet d has hinted at a $\sim$10\,amu `hybrid' atmosphere containing sulphur-bearing compounds \ce{SO2} and \ce{H2S} in a background of \ce{H2}, which is also consistent with planet d's relatively low bulk density \citep{gressier_hints_2024, banerjee_atmospheric_2024}.
\par 

The rocky planets of L 98-59 have likely experienced a similarly intense history of ionising irradiation as the inner planets of the well-known TRAPPIST-1 system \citep{agol_trappist_2021, gillon_trappist_2017}, residing in a photoevaporation regime that may inhibit atmospheric retention \citep{turbet_trappist_2020, wheatley_strong_2017, turbet_review_2020}. However, while the TRAPPIST-1 planets are roughly Earth-sized with nearly circular orbits shaped by their resonance structure, L 98-59 hosts two super-Earths, and all planets exhibit potentially large orbital eccentricities (Table \ref{tab:planets}). Compared to Earth-sized planets, secondary atmospheres on super-Earths may better withstand rapid escape under higher ionizing fluxes, due to the increasing dominance of atomic line cooling in the dynamics of more tightly bound ionospheres \citep{chatterjee2024, nakayama_escape_2022}. A recent study investigated Jeans escape of secondary atmospheres in a hypothetical L 98-59 system with b/c/d replaced by Earth-sized planets, finding that their atmospheres are stripped within 5 Gyr \citep{looveren_habitable_2025}. 

\par 
Previous estimates of the tidal heat dissipation within L 98-59 b/c/d indicate that these planets could experience significant ongoing interior heating due to their eccentric orbits \citep{quick_forecasting_2020}. The tidal heating rates provided in Table 1 of \citet{seligman_potential_2024} yield globally-averaged heat fluxes of \SI{61.1}{}, \SI{24.7}{}, and \SI{0.6}{\kilo\WPMS} for planets b/c/d respectively. Viscoelastic models of solid-body tidal deformation generally find that tidal heat flux $F_{\text{tide}}$ depends strongly on the viscosity of the planet's mantle, typically being maximised when the Maxwell time of the mantle (ratio of viscosity to shear modulus) approaches the period of tidal forcing \citep[e.g.,][]{segatz_io_1988}. For short-period rocky planets, tidal heating is maximised for a mantle viscosity $\eta \sim \SI{1e14}{\pascal\second}$ \citep{barr_tides_2018,hay_tides_2019}. 

\par 
The physical relationship between tidal heat dissipation and the rheology of the mantle raises the prospect of a self-limiting feedback. As a young rocky planet cools from its initially molten state -- and its shear viscosity increases correspondingly -- the amount of tidal heat dissipation within the solid-phase part of its mantle is expected to increase. This additional heating will slow (and potentially prevent) the solidification of the planet. An additional factor here is the presence of an overlying atmosphere, which is known to be important for controlling the cooling timescale of primordial magma oceans \citep{elkins_linked_2008, abe_early_1986, hamano_lifetime_2015, lichtenberg_vertically_2021}. Since it is primarily atmospheric composition that determines the rate at which a young planet can radiatively cool to space, it is expected that the makeup and structure of such an atmosphere also plays a role in controlling the amount of interior tidal heat dissipation. For the remainder of this paper we refer to this self-regulating feedback between radiative cooling, tidal heating, and mantle rheology as the `radiation-tide-rheology feedback'. The radiation-tide-rheology feedback mechanism modelled in this work is different to the negative feedback between tidal heating and mantle convection proposed to occur within Io and some exoplanets \citep{segatz_io_1988, wienbruch_self_1995,fischer_thermal_1990,moore_feedback_2003, matsuyama_io_2022, henning_tidally_2009}. It is also distinctly different to the `runaway melting' mechanism proposed by \citet{peale_io_1979} and \citet{seligman_potential_2024}, which works under the assumption of mantle melting from the bottom-up.

\par 
\citet{zahnle_moon_2015} previously suggested that a thick atmosphere slowed the cooling of the early Earth shortly following the Moon-forming impact \citep{canup_moon_2001}. Prolonging Earth's molten state would potentially have made tidal dissipation within its interior inefficient, significantly slowing the orbital recession of the Moon \citep{zahnle_moon_2015}. It should be noted that \citeauthor{zahnle_moon_2015} assume that a primordial atmosphere blanketing the early Earth would be composed of \SI{270}{\bar} \ce{H2O} and \SI{50}{\bar} \ce{CO2}. However, the favourable dissolution of \ce{H2O} into an underlying magma ocean makes this large partial pressure of \ce{H2O} unlikely; such an atmosphere may have been much thinner in reality, diminishing the effect of atmospheric blanketing on the early Earth. More recent modelling of tidal heating within the Hadean Earth \citep{korenaga_rapid_2023} has similarly shown the tidal heating within the planet's interior would have significantly impacted the recession of the Moon. 
\par 
The presence and strength of tidal heating within a given planet depends on the eccentricity and period of its orbit. Recently, \citet{farhat_tides_2024} developed a combined solid--fluid analytical tidal heating model. Although they did not explicitly simulate the time-evolution of any exoplanets, they found that for many of them the expected tidal heat flux is likely to exceed the incoming stellar flux, thereby yielding hotter surfaces than would be expected based on traditional radiative equilibrium calculations. They also found that the relative importance of liquid- versus solid-phase tidal heating within a planet depends strongly on its rotational period $P_d$ (equal to its orbital period if quasi-synchronously rotating) due to the different rheological behaviours of the liquid and solid phases. Fig. 4 of \citet{farhat_tides_2024} shows that liquid-phase tidal heating is minor for $P_d \gtrsim 2 \text{ days}$ and that solid-phase tidal heating is minor for $P_d \lesssim 3 \text{ days}$, though this conclusion sensitively depends on the thickness of the fluid layer \citep{hay_oceans_2019, hay_tides_2022}. However, none of these previous works self-consistently solved for coupled atmospheric energy transport (radiative transfer, convection, etc.) alongside interior tidal heating and thermal evolution.

\par 
Perhaps countering the potential for tides to keep a planet's surface molten, it was shown by \citet{selsis_cool_2023} that the presence of convectively-stable radiative layers in hypothetical pure-steam atmospheres on terrestrial planets may preclude permanent magma oceans. This is because radiative regions typically have a shallower lapse rate ($\d{T}/\d{p}$) than convective regions, allowing for radiative equilibrium to be achieved at cooler surface temperatures for a given instellation  \citep{guillot_2010, pierrehumbert_book_2010}. It was recently shown by \citet{nicholls_convective_2024} that convective stability in the atmospheres of young terrestrial planets can also be extended to mixed-gas compositions. However, \citet{nicholls_convective_2024} also found that the shutdown of atmospheric convection does not necessarily preclude the presence of an underlying magma ocean. That is, planets with radiative equilibrium temperatures less than the dry-solidus temperature can still sustain deep magma oceans due to efficient atmospheric blanketing, even for cases in which atmospheric convection has shut down. In these previous works, atmospheric convection at radiative equilibrium was triggered by the absorption of short-wave stellar radiation in the deeper atmosphere. However, atmospheric convection can also be triggered by the net upward transport of heat generated within a planet's interior. It is therefore possible that tidal heating, by its associated geothermal heat flux, could sustain atmospheric convection, thereby also influencing the atmospheric temperature structure and compositional mixing processes. 

\par 
In this work we assess how the early thermal and compositional evolution of L 98-59 b/c/d depends on the presence and magnitude of tidal heating within their interiors. Considering the complexity of the problem, two separate modelling approaches are applied:
\begin{itemize}
    \item A semi-analytic toy model (Appendix \ref{app:semi}) of planetary thermal evolution, used to probe the behaviour of the proposed radiation-tide-rheology feedback mechanism (Section \ref{ssec:res_sem}). 
    
    \item A comprehensive and physically-representative model (Section \ref{sec:methods}) of the planetary evolution, both with and without tidal heating (Sections \ref{ssec:res_noh} and \ref{ssec:res_lov}). 
\end{itemize} 
\par 
In an additionally relevant line of discussion, we use our radiative-convective atmosphere model to test the ability for the tidal heating to trigger atmospheric convection in observationally-motivated test cases (Sections \ref{ssec:res_conv} and \ref{ssec:dis_convect}). And finally, we provide crucial context for the expected atmospheric loss or retention in the L 98-59 system, drawing on models of stellar evolution and the escape of both primary and secondary atmospheres (Sections \ref{ssec:res_escape} and \ref{ssec:dis_escape}). 
\par 
Altogether, we compare the results of our modelling in Section \ref{sec:discuss} and discuss their implications for these planets in the wider context. We conclude in Section \ref{sec:conclude}.

\section{Methods}
\label{sec:methods}

\subsection{Coupled modelling framework}
\label{ssec:met_proteus}

We use the \textsc{proteus} framework \citep{lichtenberg_vertically_2021, nicholls_redox_2024} to simulate the thermal and compositional evolution of L 98-59 b/c/d. \textsc{proteus} self-consistently couples the \textsc{agni} radiative-convective atmosphere model to the \textsc{spider} interior dynamics model in order to capture the interaction between energy transport processes throughout the planet. Critically, our model also accounts for the partitioning of volatile elements between the mantle and atmospheric reservoirs. In this work, we enhance \textsc{proteus} to also include a mantle tidal heating model: \textsc{lovepy}. These modelling components are described below.

\par 
\textsc{agni} \citep{nicholls_convective_2024, nicholls_agni_2025} simulates energy transport in planetary atmospheres with an approach that preserves differentiability in the physics, allowing an optimisation method to be applied to the system (Newton-Raphson with backtracking linesearch). This numerical method determines the atmospheric temperature structure (and energy fluxes) in an energy-conserving manner, ensuring that minimal energy flux is lost across each layer up to a small numerical tolerance. Radiative transfer calculations are performed with \textsc{socrates} \citep{lacis_corrk_1991, sergeev_socrates_2023, edwards_studies_1996} using 256 \mbox{correlated-$k$} spectral bands. We include Rayleigh scattering, and band $k$-terms are fitted to line opacity data from \textsc{dace} \citep{grimm_database_2021} and continuum absorption (\citet{mlawer_mtckd_2023} and references in \citet{nicholls_redox_2024}). The \textsc{dace} database tabulates gas opacities across a wide range of temperatures and pressures, primarily drawing from the latest \textsc{exomol} and \textsc{hitemp} linelists. Dry convection is modelled with a mixing-length parametrisation under the Schwarzchild criterion \citep{vitense_mlt_1953, joyce_mlt_2023, robinson_temperature_2014}. The radiative properties of the planetary surfaces are treated with non-grey single-scattering albedos derived from lunar mare basalt \citep{hammond_reliable_2025}. As in similar studies \citep[e.g.][]{lichtenberg_vertically_2021, hamano_lifetime_2015, lebrun_thermal_2013, boer_absence_2025, elkins_linked_2008}, we treat these atmospheres as compositionally well-mixed (isochemical) and neglect disequilibrium chemical processes. Gas density is evaluated at each atmosphere level using the ideal gas equation of state according to the local temperature, pressure, and composition. The atmosphere is defined on a logarithmically-spaced grid from \SI{e-5}{\bar} to the surface pressure calculated by our outgassing scheme.

\par 
\textsc{spider} \citep{bower_linking_2019, bower_retention_2022} time-steps the interior cooling and solidification of these planets by accounting for energy transport by convection, conduction, phase separation, and gravitational settling. The interior is composed of a metallic core and a silicate mantle. The core size, defined by its radius as a fraction of the planet's interior radius, is varied in our simulation grid. The interior radius (mantle plus core) is constant throughout each individual simulation, and is calculated from the planet's dry mass (i.e. neglecting the volatile mass component) self-consistently by \textsc{spider} assuming hydrostatic equilibrium, using a pure-\ce{MgSiO3} equation of state \citep{wolf_eos_2018} for the mantle and assuming a pure-Fe core of fixed bulk density \citep[as in][slightly underestimating the bulk density for larger core radii]{bower_numerical_2018}. The dependent variables of mantle density, temperature, pressure, melt fraction, viscosity, and bulk/shear moduli all vary with radius in the modelled mantle. We assume that the mantle is initially fully-molten with an adiabatic temperature profile, consistent with previous studies of magma ocean evolution \citep[e.g][]{zahnle_moon_2015, abe_early_1986, elkins_linked_2008, hamano_lifetime_2015, krissansen_erosion_2024}. The critical melt fraction $\Phi_c$ is set to 30 per cent \citep{kervazo_solid_2021, costa_model_2009}. This value represents the point near which the material locally undergoes a rheological transition between behaving as a liquid and as a solid \citep{costa_model_2009}, and corresponds to the onset of efficient solid-phase tidal heating \citep{farhat_tides_2024, hay_tides_2019}. Radiogenic heating is neglected in this work in order to isolate the role of tidal dissipation.

\par 
\textsc{lovepy}  \citep{hay_tides_2019} calculates the tidal deformation of a spherically symmetric, radially inhomogeneous planet by numerically integrating the equations of gravito-viscoelasticity using the classic matrix propagator method \citep[e.g.,][]{sabadini_tides_2016}. The mantle is treated as a Maxwell viscoelastic solid, with melt-fraction dependent shear viscosity and shear/bulk moduli following \cite{kervazo_solid_2021}. Regions with a melt fraction above the critical disaggregation value are approximated as low shear strength solids ($\mu \sim \SI{1}{\pascal}$) \citep[e.g.,][]{bierson_io_2016}. The tidal heating rate due to shear deformation varies throughout the mantle depending on its internal structure, temperature, and rheological properties. Heating from isotropic deformation is neglected \citep{kervazo_solid_2021}. The tidal heating calculation is performed self-consistently with the time-evolution of the mantle \citep[\textsc{spider},][]{bower_numerical_2018}, which updates the depth-dependent rheological parameters as the planet's temperature structure evolves. The tidal forcing potential is due to orbital eccentricity of the synchronously rotating planet \citep{kaula_tides_1964}, and is limited to spherical harmonic degree-2. We do not include higher-order corrections in eccentricity \citep{renaud_tides_2021}, and neglect tidal-forcing imposed by one planet on another \citep{hay_tides_2019}.

\par 
Throughout the simulated planetary evolution, \textsc{proteus} determines the well-mixed outgassed atmospheric composition with eight volatile species: \ce{H2O}, \ce{H2}, \ce{CO2}, \ce{CO}, \ce{SO2}, \ce{S2}, \ce{N2}, and \ce{CH4}. Our outgassing calculation \citep{Shorttle_2024, nicholls_redox_2024, bower_linking_2019} solves for the atmospheric  partial pressures of volatiles subject to equilibrium thermochemistry \citep{JANAF} and their partitioning between the atmosphere and underlying melt, using empirically-derived solubility laws \citep{gaillard_redox_2022, sossi_solubility_2023, ARDIA201352, DASGUPTA2022291, ARMSTRONG2015283, ONEILL2002151}. As result, partial pressures depend on the total mass of melt within the planet's mantle, as well as the gravitational acceleration, temperature, and oxygen fugacity $f\ce{O2}$ at the mantle-atmosphere interface \citep{sossi_redox_2020, maurice_volatile_2024, gaillard_diverse_2021, bower_linking_2019}. See \citet{Suer2023} for a recent review on volatile solubility in magma oceans and their experimental validation. We use $\IW$ to denote the mantle oxygen fugacity, which is quantified in log-units relative to the equivalent value of $f\ce{O2}$ when set by the Iron-W\"ustite buffer at the modelled surface temperature. Magma oceans typically outgas \ce{H2}-dominated atmospheres under reducing conditions ($\lesssim\IW-2$), \ce{CO}-rich atmospheres under more moderate conditions ($\sim\IW$), \ce{CO2}-\ce{SO2} atmospheres at conditions more oxidising than $\IW+2$, and \ce{H2O}-dominated atmospheres at oxidising conditions when the mantle melt fraction is small. Appendix \ref{app:outgas} presents and discusses these different compositional regimes. 
\par 
In this work we adopt three conditions individually sufficient for model termination:
\begin{itemize}
    \item mantle solidification (where the whole-mantle melt fraction $\Phi$ is less than 2\%), 
    \item global energy balance (where the net flux transported through the atmosphere $F_\text{net}$ is equal to the tidal heat production in the interior of the planet $F_\text{tide}$, to a tolerance),
    \item a maximum integration time (200 Myr) from model initialisation.
\end{itemize}
It is important to note that we do not attempt to model the evolution of these three planets continuously up to the present day. Here, we do not self-consistently simulate atmospheric escape processes alongside their thermal evolution, nor consider differences in relative volatile inventories between the various planets.

\subsection{Planetary parameters}
Table \ref{tab:planets} outlines relevant physical parameters for these three planets in the L 98-59 system. The L 98-59 spectrum from the MUSCLES Extension is used as a template for the stellar emission \citep{behr_muscles_2023}, and is evolved self-consistently alongside the evolution of the planet using the \textsc{mors} evolution model of \citet{johnstone_active_2021}, based on the \citet{spada_radius_2013} luminosity and radius tracks. The current age of L 98-59 is estimated by \citet{engle_living_2023} to be \SI{4.94(0.28)}{\giga\year}; we adopt \SI{4.94}{\giga\year} as the current age of the star when calculating the stellar spectra used in these simulations. It has been suggested that the uncertainty on this stellar age could be larger than estimated \citep{demangeon_warm_2021}. Multiple estimates have been placed on the mass of L 98-59: $0.273\pm0.030  \text{ M}_\odot$ \citep{demangeon_warm_2021}, $0.312\pm0.031  \text{ M}_\odot$ \citep{cloutier_character_2019}, and $0.313\pm0.014 \text{ M}_\odot$ \citep{kostov_l9859_2019}. We adopt the most up-to-date estimate of $0.273 \text{ M}_\odot$ from Table A3 of \citet{demangeon_warm_2021}. Higher stellar masses correspond to larger planetary tidal heating rates, all else equal. To represent the global radiative character of a given planet with a single 1D column, we use a constant solar zenith angle of $\cos^{-1}(1/\sqrt3) = 54.74^{\circ}$ and scale the stellar spectrum by a factor of $1/4$, as per \citet{hamano_lifetime_2015} and \citet{nicholls_convective_2024}, with the assumption that all three planets are tidally locked into synchronous axial-orbital rotation.
\begin{table}
\centering
\begin{tabular}{p{0.3\linewidth} p{0.15\linewidth} p{0.15\linewidth} p{0.15\linewidth} }
\hline 
Planet                   & Planet b       & Planet c    & Planet d  \\
\hline
Mass [$M_\oplus$]        & 0.47           & 2.25        & 2.14      \\
Semi-major axis [AU]     & 0.02191        & 0.0304      & 0.0486    \\
Eccentricity             & 0.167          & 0.049       & 0.098     \\
Orbital period [days]    & 2.271          & 3.705       & 7.490     \\
\hline 
Observed radius [$R_\oplus$] & 0.850      & 1.385       & 1.521     \\
Instellation [$S_\oplus$]& 24.7           & 12.8        & 5.01      \\
Equilibrium temp. [K]    & 560.55         & 475.89      & 376.38    \\

\end{tabular}%
    \caption{Variables pertaining to the three planets considered in this work, derived from the NASA Exoplanet Archive \citep{demangeon_warm_2021, rajpaul_doppler_2024, engle_living_2023}. The first section tabulates measured quantities which are used as input parameters to our model. The second section provides additional context about other observed quantities, which are not used directly as input parameters to our models but rather calculated as dependent variables as part of our simulations. Equilibrium temperatures in this table are calculated with a Bond albedo of 30 per cent.
    }
    \label{tab:planets}
\end{table}

\par 

\subsection{Hierarchical modelling}
We consider three paradigms for modelling the role of tidal heating within the interiors of these planets. 

\par  
Firstly, we apply a semi-analytic thermal evolution model. This simplified model allows us to probe the aforementioned radiation-tide-rheology feedback, as well as potential hysteresis behaviours. It is described in Appendix \ref{app:semi} and is similar to the thermal evolution model applied in \citet{zahnle_moon_2015}.

\par 
Secondly, we apply our coupled interior-atmosphere framework \textsc{proteus} under the baseline scenario in which no tidal heating occurs. This framework is entirely distinct from the semi-analytic model. These no-tides cases provide control scenarios in which the planetary evolution pathways are primarily set by the blanketing effect from overlying atmospheres and irradiation from the star.

\par 
Thirdly, we simulate the evolution of these three planets using \textsc{proteus} while self-consistently accounting for tidal heating alongside the other aforementioned physics.

\par 
For the latter two paradigms, we run a range of simulations which vary the oxygen fugacity $f\ce{O2}$ at the magma ocean surface ($\IW-5$ to $\IW+5$) and the radius fraction $r_c$ of the metallic Fe core (50\% to 90\%) relative to the radius of the magma ocean-atmosphere interface $R_i$. In reality, $f\ce{O2}$ and $r_c$ are likely correlated, but the wide range considered for these two input variables accounts for unknown and un-modelled complexities in the redox evolution of these planets \citep[see, e.g.,][]{lichtenberg_geophysical_2022}. For comparison: Mercury has a highly reducing surface environment and a large metallic core fraction $r_c \sim 0.82$ \citep{cartier_mercury_2019, namur_mercury_2016}, while Earth has an $r_c$ of 0.55 and has had an upper mantle oxygen fugacity of approximately $\IW+4$ for at least 3.8 Gyr \citep{nicklas_redox_2018, lodders_planetary_1998, rollinson_earth_2017}. It has been suggested that the planets in the L 98-59 system have small metallic core fractions due to the host star's relatively low Fe/(Mg+Si) \citep{demangeon_warm_2021}. For the purposes of this study, a nominal radius of the rocky interior $R_i$ is approximated for each planet from its measured mass (Table \ref{tab:planets}) by \textsc{spider}, under the assumption of a hydrostatically supported pure Fe core and molten \ce{MgSiO3} mantle. This structure calculation is the same approach as in \citet{nicholls_convective_2024}, and produces a planetary structure self-consistent with our time-evolved interior modelling. These structures are not constrained to be consistent with the measured bulk densities of the planets at the present day, which are nonetheless subject to relative uncertainties of 44\% (b), 25\% (c), and 29\% (d). Uncertainties on bulk-density between these three planets arise from uncertainties on both their observed masses and radii. The initial abundances of H, C, N, and S in the magma ocean (in equilibrium with an outgassed atmosphere) are nominally set equal to estimates for the volatile composition of Earth's primitive mantle: 109.00, 109.00, 2.01, and 235.00 ppmw respectively \citep{wang_elements_2018}. The goal of this study is not to reproduce present-day atmosphere observations, but to demonstrate the effects of tidal heating coupled to magma ocean evolution. For example, while we fix these abundances to values derived from early Earth, it is known that varying the volatile inventory can strongly influence magma ocean evolution \citep{elkins_linked_2008, hamano_lifetime_2015, nicholls_redox_2024}. To this end, we perform three additional calculations of the evolution of these planets under the limiting scenario in which they have lacked atmospheres since formation.

\par 
It is known that tidal interactions can lead to changes in the orbits of planets and moons: damping eccentricity, potentially leading to orbital circularisation \citep{bolmont_tidal_2011, bolmont_formation_2014, driscoll_tidal_2015}. Continual forcing, such as in the case of mean-motion resonances between planets, acts to sustain orbital eccentricities on long timescales. This interaction is seen in the Galilean satellites \citep[e.g.,][]{peale_io_1979, yoder_tidal_1979}. We model the evolution of these planets under the assumption of orbital steady state, for simplicity. Simulations presented in Section \ref{ssec:res_ecc} address the sensitivity of their thermal evolution to different fixed values of orbital eccentricity.

\par 
Appendix \ref{app:escape} outlines our methodology for estimating the XUV irradiation and cumulative volatile losses on these three planets, assuming that it is energy-limited. Atmospheric escape is neglected within our evolutionary calculations, but it gives illuminating context to quantify the \textit{potential} for atmospheric escape on these planets. 
 
\section{Results}
\label{sec:results}

\subsection{Viscoelastic tidal heating}
\label{ssec:res_vis}

\begin{figure}
    \centering
    \includegraphics[width=0.92\linewidth, keepaspectratio]{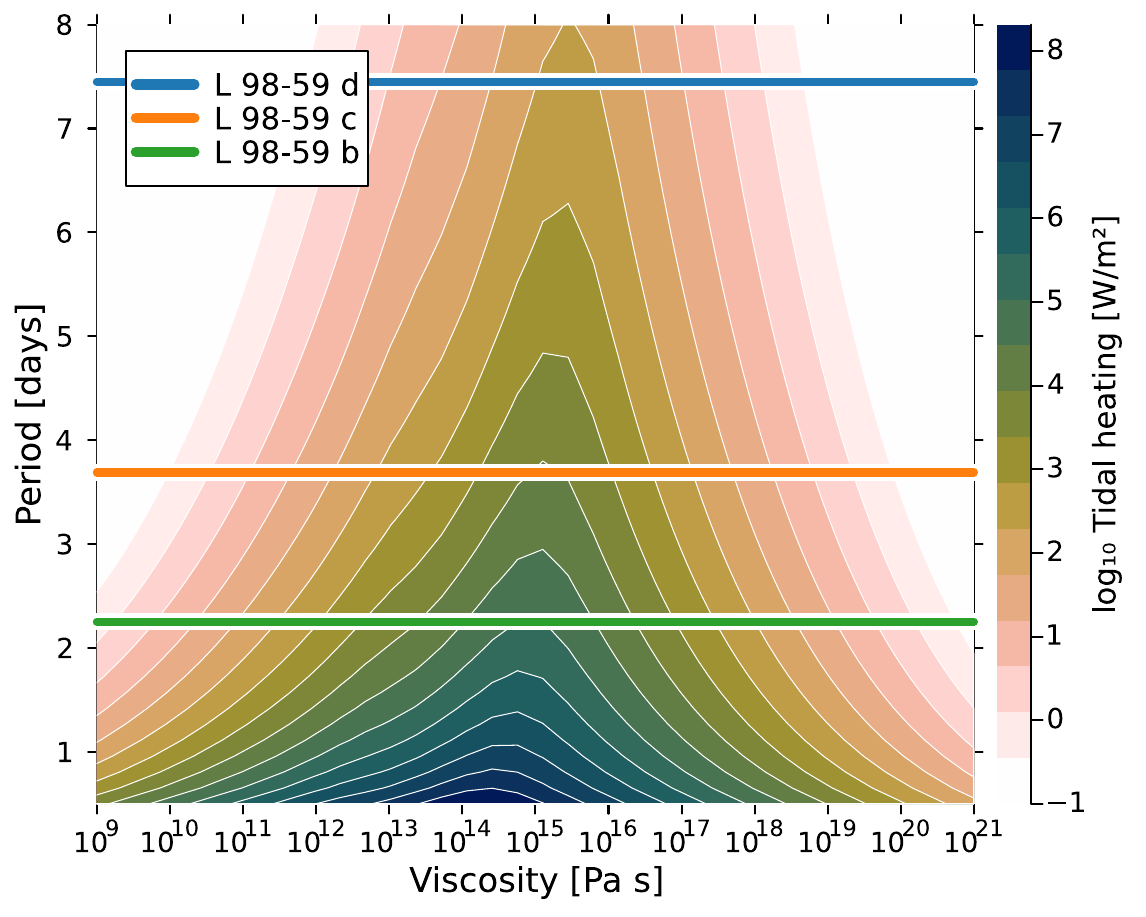}%
    \vspace*{-1mm}
    \includegraphics[width=0.90\linewidth, keepaspectratio]{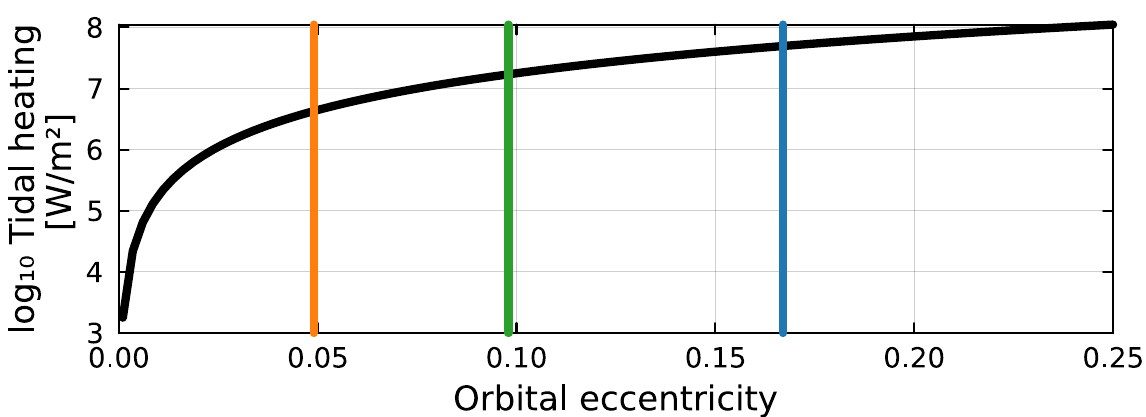}%
    \caption{Log-scaled surface tidal heat flux calculated by \textsc{lovepy} using a Maxwell viscoelastic rheology for a range of shear viscosities, orbital eccentricities, and orbital periods. For the illustrative purposes of this specific plot, the shear modulus ($\mu = \SI{60}{\GPa}$) and bulk modulus ($\kappa = \SI{200}{\GPa}$) are fixed at the solid phase end-member values from \citet{kervazo_solid_2021}. The eccentricity is set to 0.05 in the top panel. The viscosity is set to \SI{e15}{\pascal \second} in the bottom panel. Over-plotted are the estimated orbital periods and eccentricities of the L 98-59 planets (Table \ref{tab:planets}). Heat fluxes less than \SI{0.1}{\WPMS} are taken to be negligible. For comparison, Earth's current geothermal heat flux is \SI{0.07}{\WPMS} \citep{korenaga_earth_2008} and Io's is approximately \SI{2.5}{\WPMS} \citep{park_io_2024}.
    }
    \label{fig:lovepy}
\end{figure}

To first understand the expected behaviour of interior tidal heating, we use the \textsc{lovepy} \citep{hay_tides_2019} model to calculate the specific tidal heating rate for a homogeneous planet of Earth-like rheology and geometry, at various shear viscosities and orbital periods. The results of these calculations are shown in Fig. \ref{fig:lovepy}, which reproduce analytical calculations for a homogeneous body \citep{segatz_io_1988}. Heating increases for shorter orbital periods, exceeding $\SI{0.01}{\watt\per\kilo\gram}$ for periods shorter than 1 day. Of greater importance in determining the tidal heating rate is the shear viscosity $\eta$. For highly inviscid ($\eta<\SI{e9}{\pascal\second}$) and highly viscous ($\eta>\SI{e21}{\pascal\second}$) cases it is clear from Fig. \ref{fig:lovepy} that the tidal heating rate is negligible except for orbital periods shorter than 1 day. The tidal heating rate is maximised for $\eta \sim \SI{e15}{\pascal\second}$, although the exact point at which maximum heating is induced shifts to more viscous conditions for longer orbital periods as the Maxwell time increases. These behaviours compare well with previous literature \citep{driscoll_tidal_2015, henning_tidally_2009, segatz_io_1988}. 
\par 
Over-plotted on Fig. \ref{fig:lovepy} are the estimated orbital periods and eccentricities of the three L 98-59 planets of interest. It can be seen that we should expect these planets to undergo significant tidal heating when their viscosities are greater than \mbox{$\sim\SI{e9}{}$ to $\sim\SI{e12}{\pascal\second}$}. A deeper understanding of how tidal heating impacted their early evolution requires the application of our self-consistent planetary evolution model (Sections \ref{ssec:res_lov}, \ref{ssec:res_ecc}). It should also be noted that the eccentricity estimates for these planets come with significant relative uncertainties \citep{rajpaul_doppler_2024, demangeon_warm_2021}. Noting the log-scaling in the bottom panel of Fig. \ref{fig:lovepy}, it is quite possible that these planets experience different tidal heat fluxes in reality compared to those modelled here and in the literature, solely as a result of their uncertain eccentricities \citep{quick_forecasting_2020, seligman_potential_2024, farhat_tides_2024}. We test the sensitivity of our modelling to orbital eccentricity in Section \ref{ssec:res_ecc}. Furthermore, high eccentricity does not necessitate significant tidal heating, as the tidal dissipation rate also depends on the rheological behaviour of the planet's interior. For example, if the body is largely molten then tidal heat generation can be significantly reduced, potentially preserving the planet's orbital eccentricity.

\subsection{Semi-analytic modelling}
\label{ssec:res_sem}

We hypothesise (Section \ref{sec:intro}) that a negative feedback between tidal heating, mantle viscosity, and radiative cooling could extend the lifetimes of primordial magma oceans and potentially extend them indefinitely. This complex picture involves a number of potentially strong physical interactions, so here we first apply a zero-dimensional semi-analytic thermal evolution model (Appendix \ref{app:semi}). In doing so, we are not exposed to the uncertainties inherent to a more complex model, and instead aim to capture the physical interactions in a human-comprehensible manner. This particular semi-analytic modelling does not aim to make quantitative predictions of tidal heating rate and planetary evolution time-scales, but to capture the core essence of the proposed interactions and their qualitative behaviour. These calculations enable a clear interpretation of fully-coupled modelling presented in later sections. We adopt planet L 98-59 b for this baseline, taking its present-day instellation of $24.7 S_\oplus$ and a mass of $0.47 M_\oplus$ \citep{rajpaul_doppler_2024}.

\begin{figure}
    \centering
    \includegraphics[width=\linewidth, keepaspectratio]{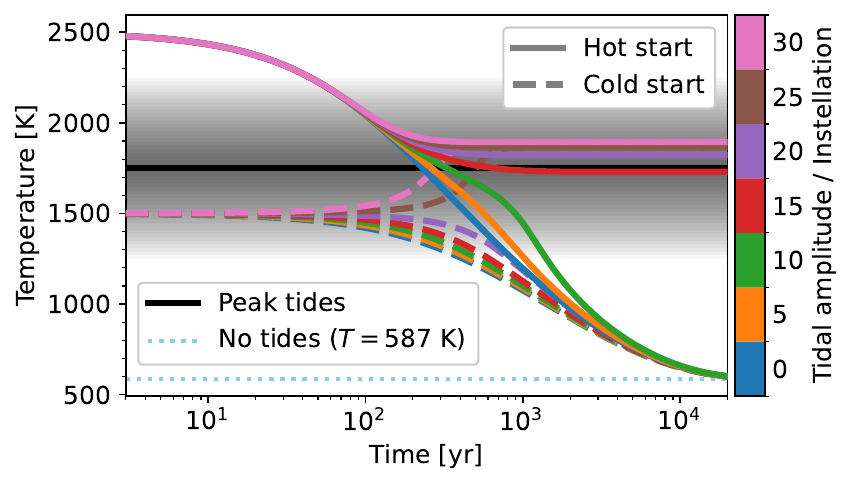}%
    \caption{Simulated behaviour for a toy analogue of L 98-59 b calculated with our semi-analytic model of thermal evolution. Temperature is plotted over time for different initial temperatures $T_0$ (line style) and tidal heating amplitudes $F_c$ (line colour). The shaded region shows the temperature regime in which tidal heating is significant, via Equation \ref{eq:semi_tides}. The black line shows the temperature at which tidal heating is maximised. The dotted blue line shows the pure-radiative equilibrium temperature.}
    \label{fig:semi}
\end{figure}
Fig. \ref{fig:semi} plots the evolution tracks for a toy analogue of L 98-59 b, calculated using our semi-analytic model (Appendix \ref{app:semi}) at various tidal heat flux amplitudes $F_c$ (line colours), and initial temperatures $T_0$ (solid/dashed line styles). In the cases of zero tidal heating (blue lines), the planet cools rapidly to its purely-radiative equilibrium temperature within \SI{20}{\kilo\year}, regardless of whether it started hot at \SI{2500}{\kelvin} or cold at \SI{1500}{\kelvin}. This short timescale is a result of the simplified interior model and lack of atmosphere. In comparison, under the most extreme case of tidal heating amplitude (pink lines) the planet sustains a high equilibrium temperature representing a permanent magma ocean. For more intermediate cases (red and purple lines) the equilibrium state of the planet is sensitive to its initial conditions. Consider $F_c = 15 S_0$: a cold-start (dashed red line) leads to an evolution pathway in which the planet solidifies without experiencing significant tidal heating, analogous to remaining on the far right of Fig. \ref{fig:lovepy}. The equivalent hot-start case (solid red line) leads to significant heating which keeps the planet in a hot state indefinitely, analogous to remaining on the far left of Fig. \ref{fig:lovepy}. A hot-start `birth' scenario is considered the typical starting point of rocky planet evolution following formation, due to the large amounts of energy delivered during accretion, by radioactive decay, and from impacts \citep{lichtenberg_review_2025, schaefer_review_2018,wyatt_debris_2008}.
\par 
Whilst the thermal evolution model applied in this section is highly simplified, it is able to generate the behaviour that we hypothesised in Section \ref{sec:intro}: tidal heating can yield a negative feedback on temperature which can prevent magma ocean solidification. How much these results extend to our more physically representative evolution model is explored in the following sections. 

\subsection{Evolution without tides}
\label{ssec:res_noh}

\begin{figure}
    \centering
    \includegraphics[width=\linewidth, keepaspectratio]{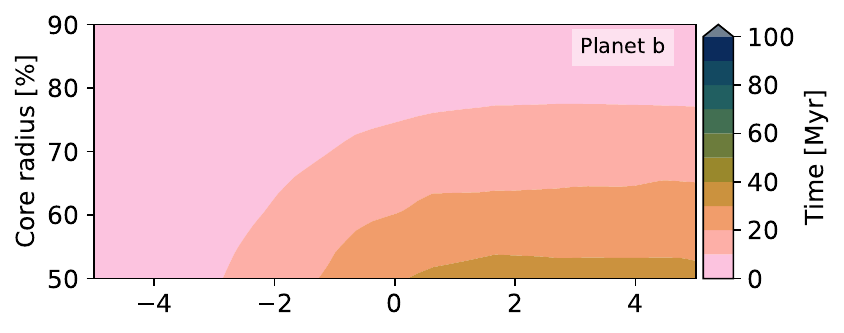}%
    \vspace*{-2mm}
    \includegraphics[width=\linewidth, keepaspectratio]{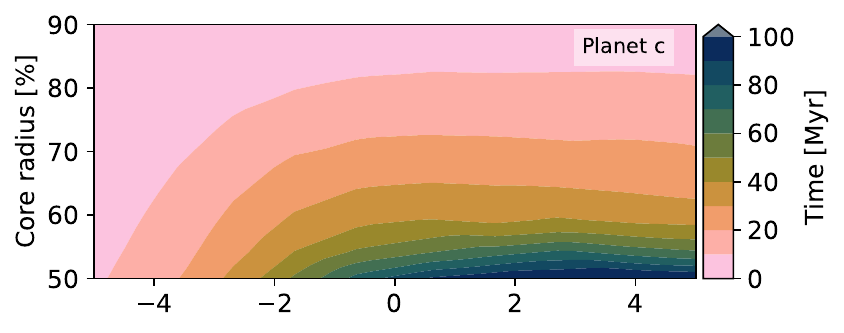}%
    \vspace*{-2mm}
    \includegraphics[width=\linewidth, keepaspectratio]{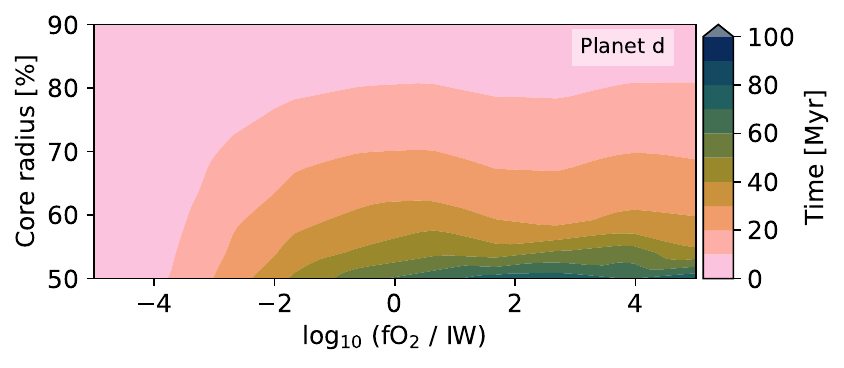}%
    \caption{Solidification time for L 98-59 b/c/d modelled by \textsc{proteus} in the absence of tidal heating. Solidification time (colour bars) varies between the planets, and also varies by metallic core radius fraction $r_c$ (y-axes) and surface $f\ce{O2}$ (x-axes).}
    \label{fig:noh_time}
\end{figure}

In this section we move beyond the semi-analytic model, and apply our full \textsc{proteus} model to simulate the time-evolution of planets b/c/d without tidal heating under different scenarios arising from accretion: varying the oxygen fugacity $f\ce{O2}$ and metallic core radius fraction $r_c$. This corresponds to 100 simulations per planet, all of which are found to result in magma ocean solidification. Following a hot-start, Fig. \ref{fig:noh_time} presents the calculated solidification time for these three planets.
\par 
All of the evolution scenarios that neglect tidal heating are found to result in mantle solidification within 100 Myr, despite blanketing by their overlying atmospheres. At all but the most reducing conditions ($f\ce{O2}>\IW-2$) these simulations also evolve close to the point of radiative equilibrium, where the net energy flux through the atmosphere is small ($|F_\text{net}| \le \SI{0.4}{\WPMS}$). Uncertainties inherent to our modelling of the stellar evolution and gas opacities could reasonably account for this small net heat flux. These uncertainties mean that, even in the absence of tidal heating, it is possible that these planets could retain shallow magma oceans for periods longer than 100 Myr. The luminosity of L 98-59 (an M3 star) will have decreased as it aged -- the evolution tracks predict that the bolometric stellar flux impinging upon these planets would decrease by $\sim20\%$ between the endpoint of our simulations up to the present day \citep{spada_radius_2013, baraffe_new_2015}. This means that even if these planets had at some point reached radiative equilibrium with a shallow magma ocean, any residual melt is likely to have solidified before the present day unless heat is continually provided by some other mechanism. We adopt a null hypothesis that the primordial magma oceans on L 98-59 b/c/d would have solidified within 100 Myr in the absence of tidal heating. 

\subsection{Evolution with tides}
\label{ssec:res_lov}

In this section we present results of modelling the evolution of L 98-59 b/c/d with \textsc{proteus} while self-consistently accounting for tidal heating within their mantles. As in \ref{ssec:res_noh} we initialise each planet into a fully-molten state, and explore a range of surface oxygen fugacities $f\ce{O2}$ and core radius fractions $r_c$. 
\par 
\begin{figure}
    \centering
    \includegraphics[width=\linewidth, keepaspectratio]{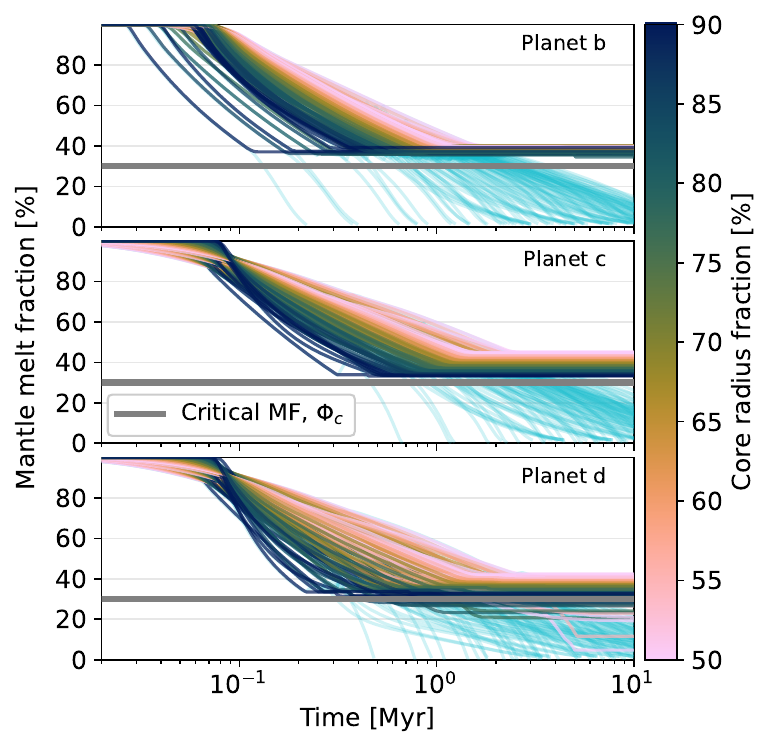}%
    \caption{Tidal heating can potentially allow L 98-59 b/c/d to maintain permanent magma oceans, given sufficient orbital eccentricty and atmospheric blanketing. Here, we plot the evolution of whole-mantle melt fraction $\Phi$ for the three planets (panels), with (colour bar lines) and without (cyan lines) tidal heating. The colour bar indicates the metallic core radius fraction $r_c$ for cases with tidal heating. The thick grey horizontal lines indicate the critical melt fraction $\Phi_c = 30\%$. We perform sensitivity tests to orbital eccentricity in Section \ref{ssec:res_ecc}.
    }
    
    \label{fig:lov_evolve}
\end{figure}
The evolution pathways for all of the \textsc{proteus} simulations with and without tides are plotted in Fig. \ref{fig:lov_evolve}. All 600 of these simulations first cool from their hot-start ($\Phi=1$, $T_\text{surf} \sim \SI{3500}{\kelvin}$), with a correspondingly large net heat flux $F_\text{net}$. The cases without tidal heating (300 cases, cyan lines) all evolve until the mantle melt fraction is effectively zero and do not achieve global energy balance (`radiative equilibrium'). Initially, tidally-heated cases evolve similarly to the no-tides cases, but their pathways diverge once efficient tidal dissipation begins. The tidally-heated cases reach global energy balance ($F_\text{tide}=F_\text{net}$) without mantle solidification, generally within 10 Myr of evolution. This difference in evolutionary outcome means that tidal heating and atmospheric blanketing are together able to slow the planets' cooling, and prevent complete solidification. The tidally-heated cases show a limiting behaviour towards equilibrium states in the vicinity of the critical melt fraction ($\Phi\gtrsim\Phi_c$), which is similar and directly analogous to the behaviour demonstrated by the semi-analytic model in Fig. \ref{fig:semi}. Atmospheric compositions depend strongly on mantle oxygen fugacity $f\ce{O2}$ \citep{seidler_impact_2024, boer_absence_2025, gaillard_redox_2022}; Appendix \ref{app:outgas} presents our calculated atmospheric compositions for tidally-heated cases at the point of global energy balance. Differences in atmospheric composition -- and correspondingly their capacity to transport energy to space -- lead to a range of evolutionary outcomes for a given $r_c$, which are visually apparent in Fig. \ref{fig:lov_evolve}. 
\par 
For planet b (top panel of Fig. \ref{fig:lov_evolve}) the melt fraction $\Phi$ does not depend strongly on $r_c$ or $f\ce{O2}$, and has a median value of $38.3\%\pm0.1\%$ at equilibrium. For planet c (middle panel of Fig. \ref{fig:lov_evolve}) the melt fraction depends primarily on $r_c$, with smaller metallic cores yielding $\Phi$ of up to 45.1\%. Due to the radiative effects of the atmospheric greenhouse -- which becomes more important at lower instellations -- the melt fraction of planet d is sensitive to $f\ce{O2}$ as well as $r_c$. It is only for planet d that $\Phi$ ever becomes less than $\Phi_c$ in these models; this planet attains the lowest simulated melt fraction across all of the tidally heated scenarios ($\Phi = 4.8\%$, for $r_c \approx 50\%$ and $f\ce{O2} \approx \IW-5$), although this region of parameter space could be less likely if $f\ce{O2}$ and $r_c$ are negatively correlated. 
\par 
Together, the results presented in Fig. \ref{fig:lov_evolve} demonstrate that tidal heating was likely sufficient to sustain the primordial magma oceans on L 98-59 b/c/d for some time after their formation. As in Section \ref{ssec:res_noh}, planets modelled with larger core radii $r_c$ generally cool faster. The median surface temperatures at the point of global energy balance in our models are: $1721.9^{+22.6}_{-23.1}$, $1771.0^{+59.4}_{-70.1}$, and $1755.4^{+47.4}_{-96.4} \text{ K}$ for planets b/c/d respectively. The ranges on these values represent the full spread of surface temperatures arising from our considerations of a range of $f\ce{O2}$ (see Appendix \ref{app:outgas}).
\par
\begin{figure}
    \centering
    \includegraphics[width=\linewidth, keepaspectratio]{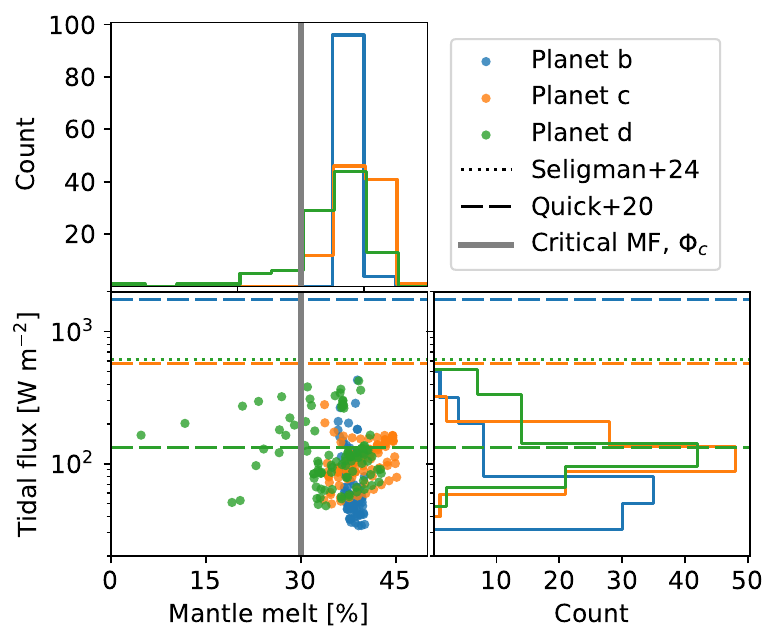}%
    \caption{Corner plot of tidal heat flux $F_{\text{tide}}$ versus mantle melt fraction $\Phi$, for tidally heated evolution outcomes of planets b/c/d (colours). This represents the point at which these simulations reach global energy balance, which is no later than 10 Myr. Dashed lines indicate tidal heat flux estimates from \citet{quick_forecasting_2020}. The dotted line indicates the heat flux estimate for planet d from \citet{seligman_potential_2024}; their estimates for planets b and c exceed \SI{2e3}{\WPMS}. The thick grey line indicates the critical melt fraction $\Phi_c$. The median melt fractions [\%] and tidal heat fluxes [\SI{}{\WPMS}] are (38.3, 52), (39.2, 101), and (36.5, 113) for planets b/c/d respectively.}
    \label{fig:lov_corner}
\end{figure}

The corner plot presented in Fig. \ref{fig:lov_corner} visualises the phase space of tidal heat flux and mantle melt fraction for the simulated outcomes of planets b/c/d. All three populations of results have unimodal distributions of mantle melt fraction (top panel), with median values greater than $\Phi_c$. The distributions of tidal heat flux (right panel) are also unimodal. The median heat flux for planet b is the smallest of the three planets, and is smaller than the predictions made by \citet{quick_forecasting_2020}. In comparison, the modelled median heat flux of planet c compares well with \citet{quick_forecasting_2020}. 

\par 
The simulations presented in Fig. \ref{fig:lov_evolve} and Fig.  \ref{fig:lov_corner} account for the greenhouse effect introduced by a range of outgassed atmospheres. For a more direct comparison with previous works \cite[e.g.][]{seligman_potential_2024}, we also ran simulations of these three planets under the limiting scenario in which they entirely lack atmospheres following their formation. Despite the loss of atmospheric blanketing, this additional scenario for L 98-59 b is still able to maintain a permanent magma ocean with a mantle melt fraction of 34.3\% and tidal heat flux of \SI{1.08e5}{\WPMS}. This is broadly comparable to the heat flux of \SI{6.11e4}{\WPMS} implicit in \citet{seligman_potential_2024}. In contrast, our simulations of planets c/d in the absence of an overlying atmosphere are found to completely solidify, resulting in small tidal heat fluxes of \SI{5.98e1}{} and \SI{4.26}{\WPMS} respectively. The stark contrast between these three no-atmosphere scenarios and those presented in Fig. \ref{fig:lov_corner} highlights the important role of the atmosphere within the radiation-tide-rheology feedback, and therefore its ability to shape the thermal evolution of rocky planets.

\subsection{Sensitivity to orbital eccentricity}
\label{ssec:res_ecc}

The power dissipated by tides within the interiors of rocky planets increases with the eccentricity $e$ of their orbits (see Fig. \ref{fig:lovepy}). However, observations of L 98-59 b/c/d have placed relatively poor constraints on their eccentricities \citep{demangeon_warm_2021, rajpaul_doppler_2024}. In this section we test the sensitivity of their thermal evolution to orbital eccentricity, by performing evolutionary calculations similar to those presented in Section \ref{ssec:res_lov}. In this section we fix $r_c = 0.55$ and $f\ce{O2} = \IW+0$, and instead vary the orbital eccentricities $e$ between 0.001 and 0.2 with logarithmic spacing. These new calculations are done with the same orbital periods as in Table \ref{tab:planets}. Models within this smaller grid of simulations are evolved past the point at which they first approach equilibrium (global energy balance), and instead terminate either at solidification or when the integration time reaches 200 Myr.
\par
\begin{figure}
    \centering
    \includegraphics[width=\linewidth, keepaspectratio]{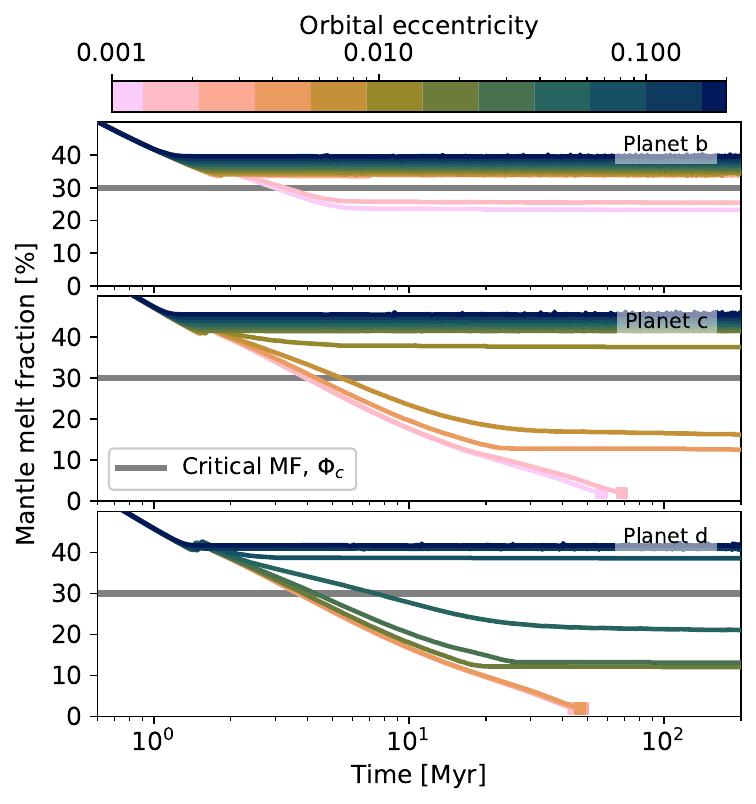}%
    \caption{Simulated mantle melt fraction (y-axes) over time (x-axis), for L 98-59 b/c/d (panels) at different orbital eccentricities (colour bar). The simulation endpoints are marked with circles. The thick grey line indicates the critical melt fraction $\Phi_c$.}
    \label{fig:ecc}
\end{figure}

 
Fig. \ref{fig:ecc} presents the results of these simulations by plotting mantle melt fraction $\Phi$ versus time. All three planets are able to achieve global energy balance with large melt fractions when the orbital eccentricity is large. Planet b (top panel of Fig. \ref{fig:ecc}) retains a large amount of melt even for relatively small eccentricities; its small orbital period enables large tidal heat production, combined with a relatively high instellation. These results show that planet b is likely to have sustained its primordial magma ocean for at least 200 Myr and potentially up to the present day. Simulations of planet c (middle panel) yield many permanently-molten cases, but are able to solidify when $e \le 0.0026$. Most of the planet d models (bottom panel) are also able to retain a large amount of melt, but cases for which $e \le 0.0069$ are able to solidify. 

\subsection{Context of atmospheric escape}
\label{ssec:res_escape}
To provide context for the atmospheric regulation of tidal heating, we analyse the history of ionizing irradiation in the L 98-59 system and constrain the resulting escape of both primary and secondary atmospheres. Our approach incorporates the \textsc{mors} model for stellar rotational evolution \citep{johnstone_active_2021} and calculates energy-limited escape rates \citep{Watson1981, Lehmer_2017}. Additionally, we consider XUV flux thresholds for the rapid escape of secondary atmospheres \citep{chatterjee2024}. Further details are provided in Appendix \ref{app:escape}.
\par
\begin{figure}
    \centering
    \includegraphics[width=\linewidth, keepaspectratio]{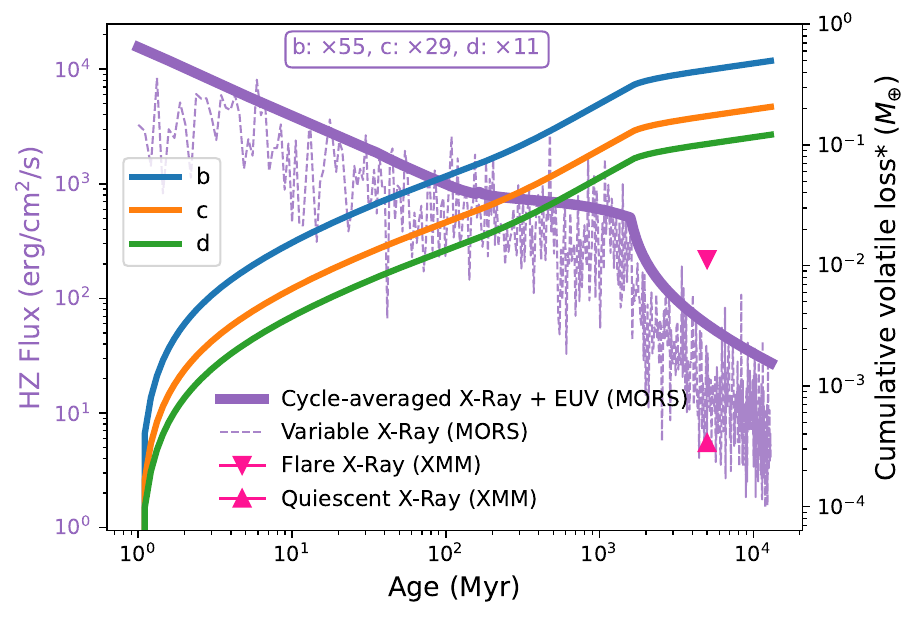}%
    \caption{Cycle-averaged XUV energy flux (solid purple line) and variable X-ray flux (dashed purple line) over time for L 98-59 ($0.273 \text{ M}_\odot$) calculated with \textsc{mors} \citep{johnstone_active_2021}, scaled to a present-day median habitable zone \citep[HZ;][]{kopparapu_habitable_2013} orbit of 0.163 AU around this star as a reference distance. Planets b/c/d respectively experience 55, 29, and 11 times more XUV radiation flux than the median habitable zone. Cumulative volatile losses for each planet (blue, orange, green lines) are plotted on the same time-axis. The cumulative volatile loss is an expression of the XUV fluence received by each planet, converted through an energy limited calculation (Equation \ref{eq:EL}).}
    \label{fig:escape}
\end{figure}

Accounting for long-term stellar variability, we show that the predicted X-ray flux displayed in Fig. \ref{fig:escape} is consistent with XMM Newton measurements of the quiescent and flare states of L 98-59 \citep{behr_muscles_2023}. The solid purple line in Fig. \ref{fig:escape} indicates that the cycle-averaged XUV flux in the habitable zone decreases from $10^{3.5}\times$ $F_{\scriptscriptstyle \mathrm{XUV}, \earth}$ at birth to $10\times$ at $5$ Gyr, where the present XUV flux at Earth is given by $F_{\mathrm{xuv}, \earth}\approx$ \SI{4e-3}{\watt \per \m \squared}. At the orbits of L 98-59 b/c/d, the XUV fluxes at $5$ Gyr are $550\times$, $300\times$ and $100\times$ $F_{\scriptscriptstyle \mathrm{XUV}, \earth}$ respectively. These levels are sufficient for any low mean-molecular-weight envelopes hosted by b/c/d to undergo rapid escape up to the present day. To illustrate the intensity of escape, we plot the potential for cumulative volatile losses for planets of b/c/d, equivalent to roughly $40\%$, $16\%$ and $10\%$ of Earth's mass over 5 Gyr (Fig. \ref{fig:escape}). These large losses mean that hydrogen and metals are likely to have escaped from L 98-59 b, while escape from planets c/d may allow the retention of metals. 
\par
The only mechanism thought to provide significant protection from photoevaporation at such high XUV fluxes is the line cooling of metal atoms, as found by \citet{nakayama_escape_2022} for an atmosphere similar to present-day Earth. This feature of cooling-limited escape could be crucial to the plausibility of atmospheric retention on super-Earths L 98-59 c/d. A comparison with the XUV threshold calculated for LHS 1140 c in \cite{chatterjee2024}, which is similar in gravitational binding to L 98-59 c/d, suggests that rapid escape of a hypothetical metal-dominated atmosphere might not occur presently while the star is within a quiescent period. However, during flaring events, the XUV flux could increase by more than an order of magnitude, potentially triggering episodic rapid escape. Previous observations of M-dwarfs have found flare duty cycles up to $25 \%$ \citep{France_2020, nicholls_temperaturechemistry_2023, venot_influence_2016}.
\par
For the sub-Venus L 98-59 b, atmospheric escape of \ce{SO2} is expected to fall between the Earth-like and Mars-like regimes explored in \citet{chatterjee2024}, where atomic line cooling fractionally reduces efficiency in the energy-limit. Consequently, at the present-day XUV flux shown in Fig. \ref{fig:escape}, the energy-limited escape of primary or secondary atmospheres will be comparable, with a shared conservative escape rate of approximately \SI{15}{\bar\per\mega\year}. For comparison, \citet{belloarufe_evidence_2025} estimate an escape rate of \SI{1}{\bar\per\mega\year} for secondary \ce{SO2} atmospheres, assuming a lower-bound 1\% efficiency.

\subsection{Sensitivity of atmospheric convection to interior heating}
\label{ssec:res_conv}
We suggested in Section \ref{sec:intro} and in \citet{nicholls_convective_2024} that interior heat production could potentially drive atmospheric convection. The \textsc{agni} atmosphere component of our modelling framework permits an assessment of this convective (in)stability. We calculate test profiles of atmospheric temperature and convective heat flux for a collection of atmospheric scenarios. 
\par 
Here, we consider scenarios in the absence of internal heating ($F_\text{net} = 0$), scenarios with a small internal heat flux ($F_\text{net} = \SI{0.1}{\WPMS}$, comparable to Earth's present surface-averaged heat flux \SI{0.07}{\WPMS} \citep{korenaga_earth_2008}), and scenarios with interior heating informed by our evolutionary calculations ($F_\text{net} = \SI{113}{\WPMS}$, the median value in Fig. \ref{fig:lov_corner}). We simultaneously consider three observationally-motivated gas compositions: \ce{H2}+25\%\ce{H2S}, \ce{H2}+13\%\ce{SO2}, and pure \ce{SO2}. The first two mixtures are consistent with the observational mean molecular weight constraints \citep{gressier_hints_2024, banerjee_atmospheric_2024}, and broadly correspond to atmospheric compositions produced by outgassing from a magma ocean containing sulfur and hydrogen. These `hybrid' compositions are formed under relatively reducing conditions (see Appendix \ref{app:outgas}). The pure-\ce{SO2} case represents an end-member which probes the pure-\ce{SO2} atmospheric compositions suggested for L 98-59 b \citep{belloarufe_evidence_2025}, although their observations do not constrain the surface pressure. The wide range of potential surface pressures ($\lesssim 30 \text{ kbar}$) probed in this additional investigation exceed those formed in the course of our main evolutionary calculations (indicated by annotations on Fig. \ref{fig:outgas}). Here, we test volatile-rich scenarios, potentially compatible with present-day observations following the planet's billion-year evolution, while our evolutionary calculations are restricted to Earth-like volatile inventories ($\lesssim 1 \text{ kbar}$) in this work.
\par 
This auxiliary investigation aims to test the energy-transporting properties of potential test atmospheres on these planets through the optically thick- and thin-regimes, in order to assess the potential for convective instability and resultant mass transport. We perform these atmospheric calculations standalone from the coupled \textsc{proteus} framework, decoupled from outgassing and interior processes. It should therefore be noted that these standalone calculations do not rely on any solubility laws (which are self-consistently applied in our main simulations; e.g. Section \ref{ssec:res_lov}), as here we focus on the atmosphere behaviours across a wide range of potential surface pressures (covering the optically-thick and -thin regimes). We do not attempt to specifically reproduce the present-day state of the L 98-59 planets, nor do we suggest that a massive \ce{SO2} atmosphere is necessarily a likely outcome from the long-term evolution of L 98-59 d. The pure-\ce{SO2} scenario is included as an end-member to probe the behaviour of \ce{SO2}.  
\par
\begin{figure}
    \centering
    \includegraphics[width=\linewidth, keepaspectratio]{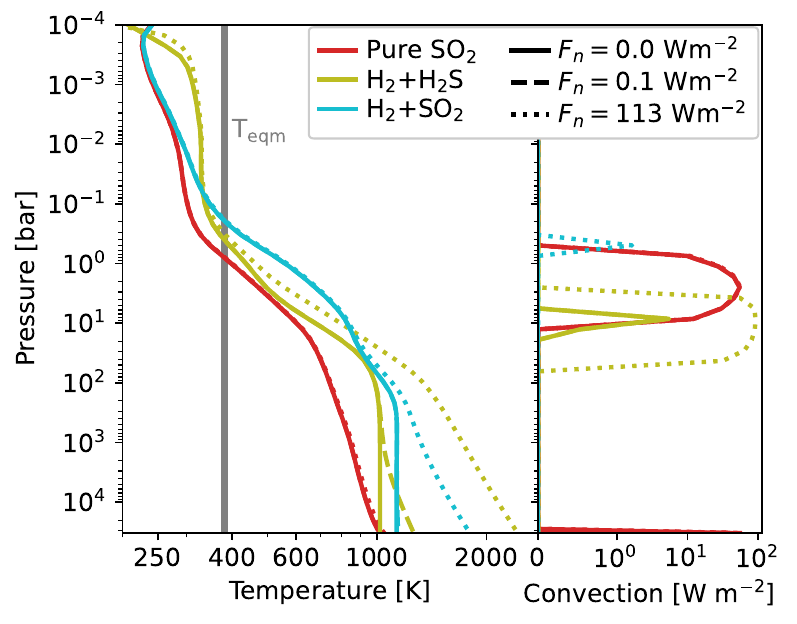}%
    \caption{Test profiles of atmospheric temperature (left) and convective heat flux (right) modelled for L 98-59 d. We consider three observationally-motivated atmospheric gas compositions (line colour) and three values for the internal heat flux (line style). Note the log scaling applied to both x-axes. The olive and cyan lines represent compositions and surface pressures (\SI{32.9}{\kilo\bar}) consistent with the retrieved mean molecular weight \citep{gressier_hints_2024, banerjee_atmospheric_2024}. The three \ce{SO2} cases (red lines) generally overlap in both panels. The vertical grey line indicates the radiative equilibrium temperature of this planet (see Table \ref{tab:planets}).}
    \label{fig:convect}
\end{figure}

 
Fig. \ref{fig:convect} shows that these example atmospheres are stable to surface-arising convection for all of the combinations of net heat flux $F_\text{net}$ and composition considered in this section. This can be seen from the right panel of Fig. \ref{fig:convect}, in which strong convection only occurs in regions with pressure $<\SI{60}{\bar}$. For the putative hybrid atmospheres (olive and cyan lines) consistent with free chemistry retrievals \citep{gressier_hints_2024}, the strength of atmospheric convection scales with the heat flux $F_\text{net}$. For a pure-\ce{SO2} composition, the convective regime is relatively insensitive to $F_\text{net}$ -- the three red profiles overlap except in the deep atmosphere. It should be noted that the pure-\ce{SO2} cases are likely much thicker than any realistic scenario for planet b, and only aim to test the radiative-convective energy transport capacity of \ce{SO2}. All cases intersect the equilibrium temperature (grey line in Fig. \ref{fig:convect}) at pressures less than \SI{1}{\bar} where the atmospheres become optically thin. The T-P ranges explored by the modelled profiles probe the vapour and supercritical regimes, and do not condense any volatiles. Although this modelling shows that reasonable values for the interior heat flux resulting from tidal dissipation (at equilibrium) are unlikely to trigger convection at the base of the atmosphere, it can nonetheless increase the middle-atmosphere convective heat flux for an \ce{H2}-rich composition. These atmospheres are able to remain convectively stable at high pressures because large temperatures make radiative diffusion efficient at transporting heat energy \citep{pierrehumbert_book_2010, nicholls_convective_2024}. It can be seen that \ce{H2}-dominated hybrid atmospheres (cyan and olive lines) are compatible with permanent magma oceans (solidus $\text{T}_\text{sol}\sim\SI{1400}{\kelvin}$; \citet{andrault_deep_2018}) when tidal heating generates a modest flux $F_\text{tide}=\SI{113}{\WPMS}$.

\section{Discussion}
\label{sec:discuss}

\subsection{Negative feedback and stable equilibria}
\label{sec:dis_feedback}
Results from our simple semi-analytic modelling of L 98-59 b (Section \ref{ssec:res_sem}) demonstrated a general negative feedback between the tidal heating of planetary interiors, their temperature-dependent rheological properties, and radiative cooling to space. Self-regulating feedbacks occurring within the interiors of rocky bodies have been previously suggested to occur within the Earth, on Io \citep{moore_feedback_2003, ojakangas_episodic_1986}, and indeed within exoplanetary interiors \citep{henning_tidally_2009}, although none of these works simultaneously addressed the physics of magma ocean evolution and atmospheric blanketing. When initialised with a hot-start, our toy semi-analytic thermal evolution models approach equilibrium with super-solidus surface temperatures given sufficiently large tidal heating. In cold-start models, tidal heating may never be sufficient to trigger a permanent magma ocean because tidal heating is less efficient at high viscosities, though more realistic rheological models \citep{bierson_rheology_2024} or eccentricity-excitation may overcome this. These results also show that limit-cycles are permissible near these stable equilibria. We have also shown that this behaviour extends to the physically-representative \textsc{proteus} models, which are initialised with a hot-start (Section \ref{ssec:res_lov}). The robustness of this feedback suggests the emergence of `tidally-supported' magma oceans within the exoplanet population: planets able to maintain magma oceans, with solidification timescales extended as a result of sustained heating by tidal interactions.

\par 
Present day planetary luminosity differences arising from cold- versus hot-start formation scenarios have been discussed within the gas-giant literature \citep{baraffe_evol_2003, marley_luminosity_2007, molliere_interp_2022}. Recent work has gone as far as to numerically constrain the entropy of the exoplanet \mbox{$\beta$ Pic b} shortly following its formation \citep{marleau_entropy_2013}. However, it is still unclear whether rocky (exo)planets form with fully molten interiors and large entropies, or with cooler interiors and correspondingly shallower magma oceans. With increasingly precise characterisations being inferred for rocky exoplanets, the radiation-tide-rheology feedback mechanism discussed in this work could be used to probe the phase space of initial conditions for a range of exoplanets (given sufficiently large orbital eccentricities). 

\par 
Abrupt transitions from one state to another, such as bifurcations, are well-studied phenomena in the context of Earth's past and current climate \citep{lorenz_flow_1963, ghil_climate_1976, steffen_traj_2018, boers_climate_2022}. In particular, there is potential for otherwise stable conditions to be disrupted by external factors such as orbital perturbations \citep{riechers_orbital_2022} and stellar evolution \citep{kasting_runaway_1988, goldblatt_runaway_2012}. Exoplanets which find themselves near stable equilibria resulting from the radiation-tide-rheology feedback (above the grey line in Fig. \ref{fig:semi}, right of it in Fig. \ref{fig:lov_corner}) could remain in this stable state indefinitely, given a fixed instellation and atmospheric greenhouse. The decreasing luminosity of M-type stars on Gyr timescales means that borderline-stable cases (e.g. red and purple lines in Fig. \ref{fig:semi}, several planet d cases in Fig. \ref{fig:lov_evolve}) could eventually undergo a `blue sky catastrophe' bifurcation leading to rapid mantle solidification \citep{meca_bsky_2004, zhou_bifur_2013}. Our results show that these stable equilibria are also in part shaped by atmospheric composition (Fig. \ref{fig:lov_corner}), so it is also possible that the loss of a greenhouse effect resulting from atmospheric escape could also induce such a bifurcation. We use our semi-analytic model to test the general feasibility of such a transition in Appendix \ref{app:bsky}, but future work should investigate the stability of these equilibria under long-term evolution.
\par 
The mass -- and therefore the luminosity -- of L 98-59 is not precisely constrained. The stellar evolution model employed in our simulations suggests that the bolometric instellation of L 98-59 b would decrease from $24.1$ $\text{ S}_\oplus$ at 100 Myr to $19.4 \text{ S}_\oplus$ at the present day (a change of $-19.5\%$). An instellation of $24.1 \text{ S}_\oplus$ is close to the median estimate of L 98-59 b's \textit{current} instellation \citep[$24.7^{+5.0}_{-4.1} \text{ S}_\oplus$,][]{demangeon_warm_2021}. The same is also true for planets c and d. Given that our estimates on the instellation of these planets are conservative, permanent magma oceans arising from our evolution models are robust to uncertainties in the current mass and luminosity of L 98-59. 
\par

In this work we have considered the time evolution of L 98-59 b/c/d. However, the modelled physics can be equally applied to other systems of rocky exoplanets. \citet{barr_tides_2018} and \citet{hay_tides_2019} found that tidal heating within the TRAPPIST-1 planets could yield large tidal heat fluxes of up to \SI{1.57e4}{\WPMS}, although these estimates were not made through an evolutionary calculation. By invoking the radiation-tide-rheology feedback process discussed in this work, it may be possible to discriminate between cold- versus hot-start formation scenarios for a range of planetary systems. Accurately characterising the tidal heating within the TRAPPIST-1 planets may also help assess the plausibility of the various atmospheric scenarios currently consistent with \textit{JWST} observations of TRAPPIST-1 b and c \citep{jkt_predictions_2022, ducrot_trappist_2024, zieba_no_2023}. For example, if an atmosphere on L 98-59 b is confirmed at high significance, while further constraints establish TRAPPIST-1 b as a bare rock, one of several explanations may be that TRAPPIST-1 b failed to reach a tidally-supported equilibrium magma ocean state, catastrophically outgassed upon early solidification, and thus exposed itself to dessication by the active young star. 
\par 

\subsection{Evolution of L 98-59 b/c/d}
\label{ssec:dis_evolve}

\subsubsection{Overview}
Turning to our coupled modelling framework \textsc{proteus}, hot-start simulations of L 98-59 b/c/d \textit{in the absence of} tidal heating all result in complete mantle solidification within 100 Myr of evolution (Section \ref{ssec:res_noh}). This behaviour is directly comparable to the solid blue/orange/green lines in Fig. \ref{fig:semi}. A cold-start for these planets would therefore also yield solidified mantles in the absence of tidal heating. The contour maps plotted in Fig. \ref{fig:noh_time} show that the solidification time, again in the absence of tidal heating, depends on both the oxygen fugacity of the mantle (through its control over atmospheric composition) and the relative size of the metallic core. By comparing the top and middle panels of Fig. \ref{fig:noh_time}, it can be seen that planet b generally solidifies more rapidly and readily than planet c, despite its higher equilibrium temperature (Table \ref{tab:planets}). These comparatively different outcomes are partially a result of the square-cube law, which allows the sub-Venus planet (b) to radiate its internal energy to space more easily than the larger planets (c and d). Planet c (middle panel) generally takes the longest time to solidify. Cases with larger $r_c$ solidify more rapidly simply due to there being a smaller amount of melt to begin with. A solidification timescale of 100 Myr in the absence of tidal heating is consistent with previous work \citep{hamano_lifetime_2015,lebrun_thermal_2013,lichtenberg_vertically_2021}.

\par 
If we instead simulate the evolution of b, c, and d self-consistently with the tidal heating model \textsc{lovepy}, it is found that all three planets are likely to have attained global energy balance whilst their mantles were still partially-molten. In any case, plotting whole-mantle melt fractions over time (Fig. \ref{fig:lov_evolve}) shows that tidal heating likely played a key role in the early evolution of the planets of L 98-59 -- and by extension, similar planets orbiting other stars -- in that their interiors are kept partially-molten for far longer than if they are not tidally heated. Long-lived magma oceans on the L 98-59 planets would become more likely considering that our simulations of the L 98-59 planets are based on conservatively-low estimates of their instellations (see Section \ref{sec:dis_feedback}). To better understand and constrain the early evolution of rocky planets and their forming atmospheres, it will be important to further observe young planetary systems \citep{Bonati2019,Cesario2024}; e.g. TOI-2076 \citep{barber_tess_2025}, TOI-1227 \citep{varga_age_2025}, and HD 63433 \citep{capistrant_hd63433d_2024}. Here, we model only the early evolution of the L 98-59 planets; if they have maintain sufficiently thick atmospheres against escape, radiation-tide-rheology feedback may potentially cause these planets to sustain magma oceans up to the present-day.

\par 
We cannot precisely predict the duration or presence of magma oceans on these planets because their initial (and current) volatile inventories are unknown. We have considered initial volatile inventories consistent with concordant estimates of volatile concentrations within Earth's primitive mantle \citep{wang_elements_2018}. Yet the lifetimes of these primordial magma oceans could potentially have been extended if the planets formed with larger initial volatile content inherited from planetary formation \citep[e.g.,][]{Venturini2020,Lichtenberg2022,Burn2024}. The outcome of magma ocean cooling depends on the planet's total hydrogen inventory due to the collisional absorption of \ce{H2} in the atmosphere and greenhouse properties of \ce{H2O} \citep{nicholls_redox_2024}. These potentially much larger envelopes would be subject to escape processes during the $\sim \SI{5}{\giga\year}$ period up to the present day. The entire hydrogen inventory of the planet need not be lost, however, given the potential of the interior as a large reservoir (multiple Earth oceans'-worth) of hydrogen, dissolved in magma \citep[e.g.,][]{DornLichtenberg2021, bower_retention_2022, sossi_solubility_2023} or solidified silicate \citep{guimond_mantle_2023}.

\subsubsection{Planet b with tides}
For the whole parameter space of $r_c$ and $f\ce{O2}$ considered in this work, the simulated mantle of planet b tends towards a narrow range of values for the mantle melt fraction ($\sim38.3\%$, top panel of Fig. \ref{fig:lov_evolve}), which is slightly larger than the adopted critical melt fraction $\Phi_c$ of 30\%. The evolutionary tracks for this planet plotted in the top panel of Fig. \ref{fig:lov_evolve} demonstrate qualitatively similar behaviour to that generated by our toy model (solid purple line in Fig. \ref{fig:semi}). The limiting feedback near $\Phi\gtrsim\Phi_c$ after an initially hot-start is key to understanding the evolutionary behaviour of planet b, and cannot be captured by the steady-state modelling of \citet{seligman_potential_2024} and \citet{quick_forecasting_2020}. The coolest surface temperature arising from tidally-heated evolutionary models of L 98-59 b is \SI{1699}{\kelvin}. This is \SI{150}{\kelvin} hotter than the $\sim \SI{1549}{\kelvin}$ surface temperature retrieved from observations by \citet{belloarufe_evidence_2025}, although they note that the surface is poorly constrained due to the upper atmosphere being near-isothermal. The equilibrium melt fractions modelled for planet b produce a unimodal distribution (blue colours in Fig. \ref{fig:lov_corner}), a direct result of the strong negative feedback between tidal heating and radiative cooling which prevents $\Phi$ from decreasing below the critical melt fraction $\Phi_c$. Results presented in the top panel of Fig. \ref{fig:ecc} indicate that even with an orbital eccentricity as low as $0.007$ \citep[consistent to $1\sigma$ with radial velocity measurements;][]{rajpaul_doppler_2024} tidal heating within this planet could prevent a primordial magma ocean from solidifying. Furthermore, an additional calculation under the limiting scenario in which planet b has always lacked an atmosphere results in a non-zero mantle melt fraction at the point of global energy balance.
\par 
A wide range of thermal evolution scenarios are consistent with L 98-59 b currently having a molten interior as a direct result of the strong radiation-tide-rheology feedback. Significant volatile loss from this planet (Fig. \ref{fig:escape}) may enable secondary eclipse observations to reliably probe its dayside temperature, and potentially constrain the amount of tidal heat production within its interior. 

\par 
Our models of planet b attain global energy balance with lower surface temperatures than those retrieved by \citet{belloarufe_evidence_2025} while simultaneously maintaining smaller tidal heat fluxes (median $F_{\text{tide}}$ of \SI{52}{\WPMS}, Fig. \ref{fig:lov_corner}) than those predicted at equilibrium by \citet{seligman_potential_2024} and \citet{quick_forecasting_2020}. A key difference between our modelling and that of \citeauthor{seligman_potential_2024} and \citeauthor{quick_forecasting_2020} is that these previous works do not account for the modulating interaction of heat transport to space introduced by an overlying atmosphere, nor the expectation that these planets are likely to have achieved global energy balance within their $\sim\SI{5}{\giga\year}$ lifetimes. Additionally, \citeauthor{seligman_potential_2024} assume that the outer lithosphere is mechanically decoupled from the interior due to an internal magma ocean. This decoupling enhances the amplitude of tidal deformation, increasing the heating rate by a factor of 2--4 for Earth-sized bodies. If we were to make an \textit{ad hoc} assumption as to the interior conditions of these planets (viscosity, density, etc.) then it is quite possible to generate predictions of large tidal heat fluxes with classic tidal theory (as solved by \textsc{lovepy}), but these are not necessarily the equilibrium states of these planets, nor representative of any late stage in their evolution. The net heat flux $F_\text{net}$ transported through the atmosphere is defined as 
\begin{equation}
    F_\text{net} = F_\text{opr} - S_0/4,
    \label{eq:radeqm}
\end{equation}
where $S_0$ is the instellation and $F_\text{opr}$ is the outgoing planetary radiation, which is itself set by the atmospheric temperature structure and composition. An atmosphere will typically produce $F_\text{opr}$ values less than that of a bare rock for a given surface temperature (Appendix \ref{app:semi}). \textit{At equilibrium}, all of the heat dissipated within the interior of the planet must be transported through the atmosphere, which requires that the tidal heat flux $F_\text{tide}$ be equal to $F_\text{net}$. Equation \ref{eq:radeqm} does not have a single solution for $F_\text{net}$ and $F_\text{opr}$ for a given $S_0$. That is, there can be multiple different interior states for a given $S_0$, with consequently different $F_\text{net}$ associated with different atmospheric compositions. This is precisely why Fig. \ref{fig:lov_corner} reveals a range of possible tidal heat fluxes and melt fractions for a given planet: we have considered a range of atmospheric compositions and temperature structures, which -- critically -- do not presume an adiabatic temperature structure nor direct cooling of the magma ocean to space. It is the combined physics of atmospheric energy transport and interior tidal heat dissipation that simultaneously determines $F_\text{tide}$, $F_\text{net}$, and the equilibrium state which makes these two fluxes equal. Our models of planet b are thereby able to achieve global energy balance with relatively cool -- although still partially molten -- interiors, and consequently mantle viscosities less than critical. These low viscosities lead to much smaller tidal heat fluxes than predicted by \citet{seligman_potential_2024} and \citet{quick_forecasting_2020}. This logic and behaviour also extends to our models of L 98-59 c and d. \citet{zahnle_moon_2015} and \citet{korenaga_rapid_2023} modelled the coupled thermal and orbital evolution of a young Earth and Moon, and also found that atmospheric blanketing of the Earth likely exerted significant control over tidal dissipation within its mantle, yielding consequently slower lunar recession.

\subsubsection{Planet c with tides}
Planet c is the most massive of the three modelled in this work, and as a result takes the longest to solidify in the absence of tides (Fig. \ref{fig:noh_time}). This proclivity for heat retention also extends to the tidally heated models (non-cyan lines, centre panel of Fig. \ref{fig:lov_evolve}) in which planet c is able to maintain mantle melt fractions of up to 45.1\% with an atmosphere, depending on the relative size $r_c$ of its metallic core. Smaller $r_c$ yield larger mantle melt fractions and a deeper magma ocean on this planet. The mantle of planet c closely approaches -- but remains strictly larger than -- the critical melt fraction (Fig. \ref{fig:lov_corner}). From a hot-start, planet c eventually tends towards a similar equilibrium state as planet b, with a partially-molten mantle. The similar evolutionary behaviour between planets b and c -- also exhibited by our semi-analytic model -- is a result of the negative radiation-tide-rheology feedback in the vicinity of $\Phi_c$ (Fig. \ref{fig:lov_evolve}), indicating that the thermal evolution of these planets is relatively less sensitive to the their particular interior structure and atmospheric chemistry compared to planet d. The thermal evolution of L 98-59 c is more sensitive to orbital eccentricity than planet b due to its longer orbital period and lower instellation; Fig. \ref{fig:ecc} indicates that eccentricities $e \le 0.0026$ may be insufficient for keeping planet c molten in the long-term. This minimum required $e$ is approximately equal to the $1\sigma$ lower limit derived from radial velocity observations \citep{rajpaul_doppler_2024}. L 98-59 c could have a permanent magma ocean to this day as a result of the radiation-tide-rheology feedback, although this scenario is likely more sensitive to the planet lacking an atmosphere than in the case of planet b.

\par 
Any observed differences in the present-day nature of planets b and c could be attributed to processes not explicitly modelled in this work, such as variable initial volatile inventories or the formation of hazes in the atmosphere of planet c. If L 98-59 b has been able to retain an atmosphere in spite of its lower escape velocity, we should therefore expect planet c to also have retained an atmosphere, all else equal. Low-confidence detections of a high molecular-weight atmosphere on planet c \citep{barclay_transmission_2023, zhou_hubble_2023} may be confirmed by future observations as part of the Hot Rocks survey \citep{diamond_hot_2023} or the STScI DDT Programme \citep{redfield_ddt_2024}. If future observations indicate that L 98-59 c lacks an atmosphere entirely, this would also point to a bare-rock explanation for L 98-59 b.

\subsubsection{Planet d with tides}
The evolutionary behaviour of planet d is the most varied of the three planets modelled in this work. Even in the absence of tidal heating, it can be seen from Fig. \ref{fig:noh_time} that its thermal evolution fundamentally depends on the oxygen fugacity of the mantle \citep[through its effect on the outgassed atmospheric composition;][]{nicholls_redox_2024}, as well as on the radius of its metallic core. Planets b and c sit within a `simpler' regime in which they are simultaneously highly irradiated (large $S_0$) and have large eccentricities (large $F_\text{tide}$), while planet d has a radiative equilibrium temperature of only \SI{376}{\kelvin}. Fig. \ref{fig:lov_evolve} shows that simulations of planet d are able to maintain a molten interior in the presence of tidal heating and an overlying atmosphere. In some cases this occurs with melt fractions $\Phi<\Phi_c$; these few cases are comparable to the solid red line in Fig. \ref{fig:semi}. Having surpassed the point of maximum tidal heat dissipation (Fig. \ref{fig:lovepy}), the green scatter points in Fig. \ref{fig:lov_corner} reveal a positive relationship between tidal heat flux $F_\text{tide}$ and mantle melt fraction $\Phi$. This relationship is simply a result of higher temperatures arising from higher internal heating rates; equivalently, that higher surface temperatures yield larger $F_\text{opr}$ all else equal. \textit{JWST} observations of L 98-59 d indicate that this planet currently has a thick volatile envelope, meaning that it has likely avoided our limiting bare-rock scenario, under which it would be expected to rapidly solidify.

\par 
Being less irradiated and having a longer orbital period, the thermal evolution of planet d is also more sensitive to its orbital eccentricity $e$. \citet{rajpaul_doppler_2024} estimate the eccentricity of L 98-59 d to be $e=0.0980^{+0.0270}_{-0.0960}$. These large uncertainties make an eccentricity as small as 0.002 consistent with radial velocity measurements ($1\sigma$). The bottom panel of Fig. \ref{fig:ecc} shows that it is possible for planet d to solidify for $e\le 0.0069$, even in the presence of tidal heating, indicating that there remains some probability that the primordial magma ocean of L 98-59 d solidified within 50 Myr. Our findings highlight the importance of placing precise estimates on the orbital parameters of rocky exoplanets, if we are to make inferences as to their past and present thermal state. 

\par 
Alternatively, a thick atmosphere on planet d could be explained by it having only recently undergone a blue sky bifurcation. Tidal heating may have historically kept its volatiles dissolved into a molten interior, protecting them from escape \citep{DornLichtenberg2021,farhat_tides_2024}, but eventually the luminosity of its star (which is not well constrained) could have decreased below the threshold for maintaining a tidally-supported magma ocean, leading to catastrophic outgassing, and resulting in a recently-formed but significant \ce{H2} envelope. This atmosphere could potentially contain \ce{H2S}, consistent with recent \textit{JWST} observations. 

\label{sec:eccentricity-spikes}
\par 
It is possible that these planets had different orbital configurations in the past compared to their present-day state. Tidally-heated planetary systems in mean-motion resonance likely undergo temporal oscillations in orbital eccentricity \citep{ojakangas_episodic_1986}. Planetary migration can also result in eccentricity amplification during a resonance crossing, inducing a spike in tidal heating \citep[e.g.,][]{cuk_dynamical_2020}. Short time-scale increases in tidal heat dissipation, such as from resonance crossings, could potentially result in episodic bursts of volcanic outgassing from a solidified night-side \citep[e.g.,][]{meier_interior_2023, boukare_deep_2022}. Modelling short time-scale processes and the complex orbital dynamics of these planets is beyond the scope of this work, however, future work should move towards coupling planetary thermal and orbital evolution. We have shown that time-dependent processes with hysteresis play an important role in setting the present-day state of these planets. Upcoming future observations may attempt to place limiting constraints on these planets' historical orbital states: for example, the presence of a present-day magma ocean on L 98-59 d could potentially suggest a historically larger orbital eccentricity, inferred via evolutionary modelling as in this work.

\subsection{Long-term atmospheric replenishment}
\label{ssec:dis_escape}

Following from the updated limits to escape this study has provided (Section \ref{ssec:res_escape}), and their ensuing constraints on volatile mass-balance, we can re-address what is implied by the detection of any atmosphere today. There are two limits to atmospheric revival: (\textit{i}) the current outgassing rate must be at least as high as the current escape rate, and (\textit{ii}) the initial volatile inventory must be at least as high as the actual accumulated volatile loss. 

\par 
The first condition, applied to L 98-59 b, states that the current volatile outgassing rate must be at least $\sim$4.5$\times 10^6$\,kg\,s$^{-1}$ to sustain an atmosphere \citep[Fig. \ref{fig:escape}; tenfold greater than the lower limit of][]{belloarufe_evidence_2025}. Is this outgassing rate plausible? The methods of the present study cannot answer this question directly because we have modelled magma ocean degassing as a vapour-pressure equilibrium, assuming instantaneous volatile transport within the magma ocean itself and no escape to space; the degassed partial pressure of each species in our model is set by empirically-derived solubility laws and chemical equilibrium. Magma ocean degassing may be limited by diffusion across the upper thermal boundary layer \citep[see Equation S7 in][]{hamano_emergence_2013}. However, the very high surface heat flux and thin thermal boundary layer in our models would point to fast diffusion. Alternatively, we can consider the case of a mostly-solidified planet outgassing in a heat pipe mode, as in \citet{belloarufe_evidence_2025}. In this case, the equilibrium outgassing rate would be related to tidal heating via the energy-balance assumption that all interior heat is transported by melting and melt advection \citep{oreilly_magma_1981, moore_io_2001}. A conservative surface temperature of \SI{800}{\kelvin} and a tidal heat flux $F_\text{tide}$ of \SI{50}{\WPMS} (Fig.  \ref{fig:lov_corner}) permits an equilibrium extrusive volcanism rate (resurfacing rate) on the order of $10^{10}$\,kg\,s$^{-1}$ for planet b \citep{oreilly_magma_1981}. The outgassing rate required to balance escape of $\sim$4.5$\times 10^6$\,kg\,s$^{-1}$ \ce{SO2} could be sustained if the magma contained \ce{SO2} at a respectable concentration of $\sim$200\,ppm, assuming all \ce{SO2} degasses completely. However, integrating this magma production rate back in time over 1\,Gyr indicates that this melting-degassing cycle would have processed 200 planets'-worth of mantle rock. It is not obvious whether much sulphur would remain in the mantle residue to be outgassed in the future after so many melting cycles, unless the outgassed sulphur is easily reburied into the mantle \citep[as on Io;][]{dekleer_isotopic_2024} rather than being lost to space.

\par 
Assessing the second condition -- the `supply limit' -- is less straightforward because an atmosphere cannot be lost to space if it does not exist. One might imagine a scenario where several Earth ocean masses of water are sequestered within a planet's solidified interior, and the planet has been in a tectonic regime that does not permit efficient volcanic outgassing \citep[such as a stagnant lid regime;][]{guimond_low_2021}. This scenario would mean that the actual escape rates are lower than theoretically possible for some time, due to a limit on atmospheric supply. A sudden increase in volcanic activity \citep[e.g. a tectonic regime shift;][]{lenardic_solar_2016} would replenish an atmosphere in spite of its high potential to have escaped in the past \citep{kite_atmosphere_2020}. 
One potential way to effect episodes of strong volcanic outgassing is through exciting the eccentricity of the planet (e.g., in a resonance crossing event; see Section \ref{sec:eccentricity-spikes}), which would temporarily increase the tidal heating rate, melting parts of the otherwise largely solid mantle. In any case, the potential 5 Gyr cumulative volatile loss on L 98-59 b is $\gtrsim$130\% of its current mass (Figure \ref{fig:escape}), implying that if an atmosphere is present today, its outgassing rate must have been much lower in a previous era. 

\par 
Overall, it should be noted that whether prolonged magma oceans help or hinder atmospheric retention in the long term remains an open question. Magma oceans can hold large amounts of volatiles \citep{DornLichtenberg2021, nicholls_redox_2024, sossi_solubility_2023}. For a given inventory of volatiles, dissolution into an underlying magma ocean would yield thinner atmospheres with smaller radii, thereby potentially decreasing the escape rate \citep[e.g.,][]{maurice_volatile_2024}. Alternatively, early magma ocean solidification could trap volatiles within the \textit{solid} silicate \citep{tikoo_fate_2017, hier_origin_2017, sim_volatile_2024}. Although mantle minerals certainly have lower solubility limits for volatile species compared with magma, a solid mantle is, unlike surface magma, an inefficient degasser: longer residence times in the solid mantle offer a distinct refuge against atmospheric exposure \citep{kite_atmosphere_2020}. Towards the provisioning of S-rich atmospheres in particular, the trade-off is even more unclear, given that sulphur could be stored in the mantle in excess of saturation, as accessory phases \citep[e.g.,][]{boukare_production_2019, lark_sulfides_2022, guimond_stars_2024}. A more thorough understanding of how magma oceans affect volatile retention, by applying coupled outgassing-escape models, will be key to knowing how readily rocky planets retain atmospheres.

\subsection{Atmospheric convection, mixing, and structure}
\label{ssec:dis_convect}

It has been previously shown that outgassed secondary atmospheres at pure radiative equilibrium (net heat flux $F_\text{net}=0$) exhibit deep radiative layers which are stable to convection \citep{selsis_cool_2023, nicholls_convective_2024}. In addition to short-wave heating, atmospheric convection can be sustained by ongoing heating within a planet's interior; this is equivalent to $T_{\text{int}}>0$ in the nomenclature of the gas-giant literature \citep[e.g.][]{parmentier_nongrey_2015}. Inferences of \ce{SO2} and \ce{H2S} in the upper atmosphere of L 98-59 d have been suggested to be caused by surface volcanism \citep{gressier_hints_2024}, necessitating upward transport of gas to the level probed by transmission spectroscopy. While all of our simulations assume isochemical atmospheres, Section \ref{ssec:res_conv} provides context and motivation for future work by assessing the potential for convective transport in test atmospheres. We emphasise that the atmospheres modelled in Fig. \ref{fig:convect} are decoupled from interior interactions, and only test the atmospheric energy-transport for a handful of compositional scenarios relevant to recent observations of L 98-59 d, in the context of potential tidal heating.
\par 
Fig. \ref{fig:convect} shows that interior heating -- such as by tidal dissipation -- can trigger and/or strengthen the convection in the atmosphere of L 98-59 d. Under atmospheric scenarios consistent with the retrievals \citep{gressier_hints_2024, banerjee_atmospheric_2024}, the temperature and convection profiles within optically thick regions are sensitive to $F_\text{net}$. For the \ce{H2}+\ce{H2S} cases (olive lines), a small internal heat flux of \SI{0.1}{\WPMS} is able to raise the surface temperature by \SI{244}{\kelvin}. A larger heat flux (olive dotted line) informed by our tidal heating models raises the surface temperature by $+\SI{1421}{\kelvin}$. Both of the \ce{H2}-dominated cases modelled here maintain surface temperatures greater than a representative solidus temperature \citep{bower_linking_2019}, consistent with a permanent magma ocean. While \ce{SO2} has unique spectral features in the infrared that have been leveraged by recent observations of these planets \citep{belloarufe_evidence_2025}, its broadband opacity is relatively low \citep{underwood_so2_2016}. This means that the atmospheres rich in \ce{SO2} presented in Fig. \ref{fig:convect} are comparatively less sensitive to $F_\text{net}$, and yield systematically lower surface temperatures than the other compositions considered. The temperature profiles of the red and cyan lines converge in the upper atmosphere, where the opacity of \ce{H2} becomes negligible and \ce{SO2} remains the primary absorber of radiation.
\par 
The strength of atmospheric convection increases with $F_\text{net}$, which would also act to increase compositional mixing for this particular atmospheric composition. However, it is still unclear whether tidally-enabled atmospheric convection could lift species outgassed at the surface to the observed level, as the convection does not extend immediately from the surface upwards. Other mixing processes -- Rossby waves, gravity waves, synoptic eddies, molecular diffusion, etc. -- will need to be invoked to explain the transport of outgassed volatiles to the observable regions of these atmospheres \citep{pierrehumbert_book_2010}. \textit{In situ} photochemical production may explain tentative detections of \ce{SO2}, since the large molecular weight of this molecule combined with a shallow lapse rate (Fig. \ref{fig:convect}) will make its diffusion into the upper atmosphere highly inefficient \citep{tsai_photochemically_2023, seinfeld_atmospheric_2006}.  As a third alternative to \textit{in situ} high-altitude \ce{SO2} production or passive mixing processes, we note that explosive volcanic plumes have been suggested to loft \ce{SO2} into the upper atmosphere of Venus \citep[e.g.,][]{esposito_sulfur_1984}, which, if identified as a source (i.e., based on their transient nature), would rule against magma ocean degassing. However, it is not clear whether even such explosive plumes would remain buoyant to high altitude in a thick hydrogen-rich atmosphere \citep[cf.][]{mastin_userfriendly_2007}.

\par 
We treat these planets with a 1D model. This simplification is frequently adopted across the related literature for the purposes of computational feasibility, given flexible and accurate parametrisations. Spatial variations in tidal heat flux with latitude, although likely no more than a factor of 2--4 \citep{beuthe_tides_2013, hay_oceans_2019}, could potentially drive local atmospheric dynamics \citep{wallace_dynamics_2006, lemasquerier_europa_2023}.

\par 
Gas opacity under extreme conditions is poorly understood due to computational limitations on compiling linelists for certain molecules and experimental limitations when empirically deriving absorption cross-section databases \citep{grimm_database_2021, mlawer_mtckd_2023, BIRK201788, komasa_h2_2011}. Potential inaccuracy in gas absorption properties (line and continuum) may systematically impact conclusions across the exoplanetary literature, including this work, although here we have drawn from contemporary databases designed for modelled high-temperature regimes \citep{nicholls_redox_2024, nicholls_agni_2025}. Furthermore, uncertainties in gas thermodynamics (e.g. their equations of state) at high pressures will bias modelling results  \citep{wordsworth_review_2022,haldemann_biceps_2024}. In this work we adopt isochemical atmospheric profiles and the ideal gas equation of state -- since we do not focus our investigation on atmospheric structure -- which are assumptions broadly applied in previous studies (e.g. \citet{Cherubim_2025, wordsworth_2013, schlichting_chemical_2022, Katyal_2019, rogers_road_2025, Krissansen-Totton2024Sep}). In a follow-up work we include more realistic gas-phase equations of state into our \textsc{agni} atmosphere model, since non-ideal behaviours may impact inferences made from observations of exoplanetary radii and the location of convective zones. Lab experiments on these processes, and particularly on volatile solubility into planetary interiors \citep{ANZURES202561, lichtenberg_review_2025}, are increasingly necessary for developing sufficiently comprehensive models to explain observations of exoplanets.

\section{Conclusions}
\label{sec:conclude}

We have used a coupled atmosphere-interior modelling framework to simulate the early evolution of the sub-Venus L 98-59 b and the super-Earths L 98-59 c/d. Simulations with a self-consistent implementation of tidal heating, mantle rheology, and atmospheric energy transport indicate that the primordial magma oceans of these three planets could have been sustained for some time. This is possible, despite their sub-solidus radiative equilibrium temperatures, due to a robust negative feedback mechanism between tidal heating, mantle rheology, and radiative cooling. Our conclusions are summarised as follows:

\begin{enumerate}
    \item The `radiation-tide-rheology feedback' reported here is central to controlling the early evolution of rocky planets, as it introduces stable equilibria which can indefinitely prevent their interiors from completely solidifying.

    \item Solid-phase tidal heating played a key role in the early thermal evolution of L 98-59 b/c/d, and likely prolonged the lifetimes of their primordial magma oceans. Across a wide parameter space, our models find that L 98-59 b may have permanent magma ocean to this day due to the proposed feedback mechanism, whether this planet has retained an atmosphere or not. Permanent magma oceans could also extend to planets c and d for as long as they retain atmospheres, subject to other processes (such as atmospheric escape).

    \item When considering tidal heating within a coupled framework, atmospheric energy regulation reduces planetary tidal heat fluxes by up to two orders of magnitude compared to previous estimates. We emphasise that coupled models are key to understanding the behaviour of these non-linear systems, and that future work should include coupling with atmospheric escape processes.
    
    \item An atmosphere on the sub-Venus L 98-59 b would be in a regime of rapid photoevaporation, whether primary or secondary, throughout its history. This is in agreement with previous studies. Indications of an atmosphere on this planet would require further theoretical explorations of its thermal and compositional evolution.  
    
\end{enumerate}
 
Further theoretical and observational studies of the L 98-59 system will enable us to test the coupled physics of magma ocean thermal evolution, mantle redox chemistry, atmosphere-interior interactions, and escape processes. Our simulations indicate that L 98-59 b could very well be a `tidally-supported world'. However, future theoretical studies should attempt to model the complete evolution of these planets with a self-consistent representation of atmospheric escape and longer term stellar evolution. L 98-59 b/c/d are `targets under consideration' as part of the upcoming STScI DDT programme \citep{redfield_ddt_2024}. Secondary eclipse observations will constrain their dayside surface temperatures, and allow inferences as to whether they are currently molten or not. L 98-59 c and d will also be observed with \textit{Pandora}, which is scheduled to launch at the end of 2025 \citep{barclay_pandora_2025}.

\par 
Future studies on the role of tides on exoplanet evolution should account for changes in their orbital parameters (eccentricity, instellation, period) over time \citep{ojakangas_episodic_1986, bolmont_tidal_2011, bolmont_formation_2014, heller_tidal_2011}. Furthermore, more appropriate rheological models should be employed \citep{castillo-rogez_tides_2011,renaud_rheology_2017, bierson_rheology_2024}, and dissipation in the fluid should be accounted for \citep{farhat_tides_2024}, particularly in tightly-packed planetary systems \citep{hay_io_2020}. We also highlight the necessity of precise constraints on orbital parameters, if we are to make inferences as to the past and present thermal state of rocky exoplanets.

\par 
Based on recent measurements by the JUNO spacecraft, \citet{park_io_2024} find that Io does not contain an internal magma ocean and instead has a mostly solid mantle. Based on this, \citeauthor{park_io_2024} suggest that strong tidal heating within planetary interiors may not always be sufficient to \textit{create} magma oceans within planetary interiors. However, the physics of the radiation-tide-rheology feedback reported here suggests that, alternatively, primordial magma oceans may not cool in the first place; long-lived magma oceans could still be common on rocky planets with short period eccentric orbits.


\section*{Acknowledgements}
CRediT author statements. \textbf{HN}: Conceptualization, Methodology, Software, Formal analysis, Investigation, Validation, Writing - Original Draft, Writing - Review \& Editing, Visualization. \textbf{CMG}: Conceptualization,  Writing - Original Draft, Investigation, Validation. \textbf{HH}: Conceptualization,  Methodology, Writing - Original Draft, Investigation, Validation. \textbf{TL}: Conceptualization, Methodology, Supervision, Resources, Writing - Review \& Editing. \textbf{RDC}: Methodology, Formal analysis, Software, Conceptualization, Writing - Original Draft, Investigation, Visualization. \textbf{RTP}: Conceptualization, Supervision, Writing - Review \& Editing. 
\par 
TL acknowledges support from the Netherlands eScience Center under grant number NLESC.OEC.2023.017, the Branco Weiss Foundation, the Alfred P. Sloan Foundation (AEThER project, G202114194), and NASA’s Nexus for Exoplanet System Science research coordination network (Alien Earths project, 80NSSC21K0593). CMG is supported by the UK STFC [grant number ST/W000903/1]. HCFCH is supported by the Leverhulme Trust Research Project Grant (RPG‐2021‐199). RTP and RDC acknowledge support from the UK STFC and the AEThER project. 
\par 
We thank our two anonymous reviewers for their encouraging feedback and constructive suggestions.

\section*{Data Availability}
\textsc{proteus}\footnote{\url{https://github.com/FormingWorlds/PROTEUS}} and  \textsc{agni}\footnote{\url{https://github.com/nichollsh/AGNI}} are open source software available on GitHub. The version of \textsc{lovepy} used in this article is archived on Zenodo\footnote{\url{https://doi.org/10.5281/zenodo.15610940}}. Where possible data underlying this article are available on Zenodo, at \url{https://dx.doi.org/10.5281/zenodo.14832151}. The remaining large data files are available upon request to the authors.


\bibliographystyle{mnras}
\bibliography{main.bib}


\appendix

\section{Semi-analytic evolution model}
\label{app:semi}
In this appendix we derive a simple semi-analytic planetary evolution model. This model aims to capture the general physical interactions occurring throughout these planets in an easily comprehensible manner, without being exposed to the uncertainties which naturally arise from our more complex model (Section \ref{ssec:met_proteus}).  Here, the planet is considered a homogeneous sphere with average density ($\rho = \SI{4550}{\kilo\gram\per\meter\cubed}$), heat capacity ($c_p = \SI{1250}{\joule\per\kilo\gram\per\kelvin}$), solidus ($T_s = \SI{1500}{\kelvin}$), and liquidus ($T_l = \SI{2000}{\kelvin}$). These values are taken to be broadly consistent with the properties of Earth's mantle \citep{stixrude_thermo_2005}. Given a mass $m$, the radius can be simply calculated as
\begin{equation}
    R = (3 m/4 \pi \rho)^{1/3}.
\end{equation}
\par
The short-wave surface albedo $\alpha$ is set to 0.2 and the long-wave emissivity $\varepsilon$ is set to unity \citep{fortin_lava_2024, hammond_reliable_2025}. The planet is irradiated on its day-side with a bolometric flux $S_0$, and its surface emits radiation directly into space isotropically as a greybody according to the Stefan-Boltzmann law
\begin{equation}
    F_u = \sigma \varepsilon T^4.
\end{equation}
The net power $P_n$ entering the planet as a function of its temperature $T$ can then be expressed as an energy balance between the absorbed, internal, and emitted fluxes:
\begin{equation}
    P_n(T) = \pi R^2 (1-\alpha) S_0 + 4 \pi R^2 F_i  - 4 \pi R^2 \sigma \varepsilon T^4,
    \label{eq:semi_pow}
\end{equation}
where $F_i$ is the internal heat flux (such as from tidal or radiogenic heating). This expression for the thermal evolution of the planet is identical to the interior evolution model of the young Earth set out by \citet{zahnle_moon_2015}. Through the chain rule, Equation \ref{eq:semi_pow} can be converted to a heating rate
\begin{equation}
    \frac{dT}{dt} = \frac{4 \pi R^2}{m c_p} \bigg( \frac{1-\alpha}{4}S_0 + F_i - \sigma \varepsilon T^4 \bigg).
    \label{eq:dTdt}
\end{equation}

\par 
If we were to take $F_i=0$, the solution to equation \ref{eq:dTdt} is that of planetary radiative equilibrium. However, the inclusion of $F_i$ as a function of temperature $T$ permits a semi-analytic evolution model which parametrises the effects of tidal heat dissipation. Maxwellian models of tidal heat dissipation demonstrated by \textsc{lovepy} (Fig. \ref{fig:lovepy}; \citet{hay_tides_2019}) and other works \citep{driscoll_tidal_2015, henning_tidally_2009} generally yield a unimodal relationship between the amount of heat dissipated and the mantle melt fraction. For the purposes of our simplified semi-analytic modelling, we heuristically parametrise tidal heating with a Gaussian function
\begin{equation}
    F_i(T) = F_c \exp\bigg[ -\bigg(\frac{T-T_c}{T_w}\bigg)^2 \bigg],
    \label{eq:semi_tides}
\end{equation}
where $T_w$ characterises the width of the temperature range sensitive to heating, $T_c$ is the temperature at which tidal heating is maximised, and $F_c$ is the amplitude of the tidal heating. We centre this heating function between the solidus and liquidus of the material. In reality, the amount of heat dissipated by tidal stresses depends on a number of factors which are robustly accounted-for in our \textsc{proteus} simulations (Section \ref{ssec:met_proteus}). A Gaussian function is an appropriate choice for the level of complexity desired of this semi-analytic model, as it has a single global maximum which decreases monotonically to zero at $\pm\infty$ (qualitatively representing a negligible amount of heat dissipated by tides in the fully-solid and fully-molten regimes).
\par 
These simplifications allow us to express the complete form of our semi-analytic evolution model as a single equation:
\begin{equation}
    \frac{dT}{dt} = 
        \frac{4 \pi R^2}{m c_p} 
        \bigg( \frac{1-\alpha}{4}S_0 +
                F_c \exp\bigg[ -\bigg(\frac{T-T_c}{T_w}\bigg)^2 \bigg] - 
                \sigma \varepsilon T^4 
        \bigg).
    \label{eq:semi}
\end{equation}
which is nominally integrated over time from $(T,t) = (T_0,0)$ up to sufficiently large limit of $t=\SI{20}{\kilo\year}$, which provides sufficient time for the modelled cases to reach steady-state (see Fig. \ref{fig:semi}).

\section{Blue sky bifurcations arising from stellar evolution}
\label{app:bsky}

In Section \ref{sec:dis_feedback} we suggest that planets which find themselves oscillating in equilibria through radiation-tide-rheology feedback could potentially still solidify. This could occur for situations in which tidal heating is initially sufficient to sustain a permanent magma ocean, but only in combination with intense stellar irradiation. The slowly decreasing luminosity of an M-type host star raises the possibility of a planet sustaining a magma ocean for several Gyr, before rapidly solidifying through a `blue sky catastrophe' \citep{meca_bsky_2004, zhou_bifur_2013}.
\par
\begin{figure}
    \centering
    \includegraphics[width=\linewidth, keepaspectratio]{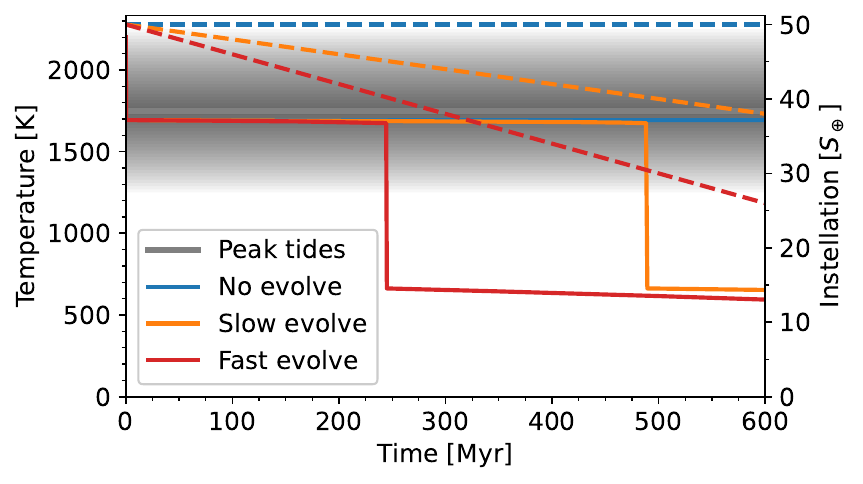}%
    \caption{Thermal evolution of a toy planet subject to tidal heating with a co-evolving star. Solid lines plot the planet's temperature over time for three cases of stellar evolution (line colours). Dashed lines plot the correspondingly evolving instellations. The thick black line indicates the temperature at which tidal heating is maximised.}
    \label{fig:bsky}
\end{figure}

 
In this appendix we apply the semi-analytic toy model described in Appendix \ref{app:semi} to a test planet with a fixed arbitrary $F_c$. This is done under the consideration of three scenarios for how $S_0$ could change over time: constant, slowly decreasing, and rapidly decreasing. In reality, bolometric stellar luminosity does not decrease strictly linearly in time \citep{baraffe_new_2015}. Fig. \ref{fig:bsky} presents these three additional simulations, with the different stellar evolution scenarios indicated by the line colours. The case without stellar evolution (blue lines) remains in radiation-tide-rheology equilibrium with a hot surface. The cases in which $S_0$ decreases undergo a bifurcation when the $S_0$ drops below a critical value (approximately 80\% of its initial value in this case). This bifurcation represents the point at which tidal heating is unable to sustain equilibrium (global energy balance), causing the planet to rapidly cool to a temperature which can be sustained by the instellation alone. 

The particular values of $S_0$ considered in this section are not important because we consider here a toy thermal evolution model (Appendix \ref{app:semi}) in which $S_0$ and the albedo $\alpha$ are effectively degenerate in forcing the thermal evolution of the planet. Any decrease to $S_0$ could be compensated by some  decrease to $\alpha$; quantitatively modelling potential planetary evolution pathways necessitates a comprehensive modelling tool such as \textsc{proteus} (Section \ref{ssec:met_proteus}). Fig. \ref{fig:bsky} shows that it is, however, still important to consider that rapid changes to a planet's thermal state may occur later in its lifetime, potentially driven by slow changes to the amount of radiation recieved from its host star.

\par
\begin{figure}
    \centering
    \includegraphics[width=\linewidth, keepaspectratio]{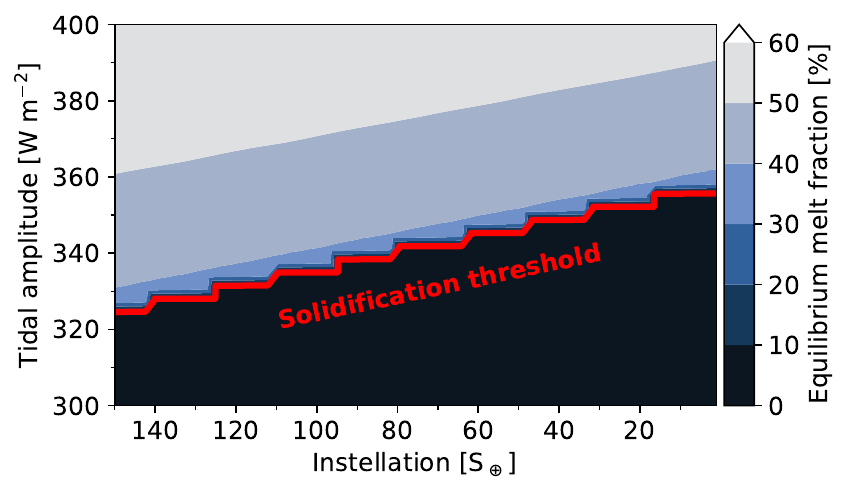}%
    \caption{Equilibrium melt fractions $\Phi$ for a toy planet calculated with our semi-analytic thermal evolution model, for a wide range of instellations $S_0$ and tidal heating amplitudes $F_c$. The red contour line divides the space at $\Phi=1\%$.}
    \label{fig:bithresh}
\end{figure}

 
Fig. \ref{fig:bsky} reveals that bifurcations to a cool state may indeed be triggered once the instellation drops below some critical value. To elucidate how this compares against the strength of our (parameterised) tidal heating within the semi-analytic model, we simulated the evolution of the toy planet under different tidal heating amplitudes $F_c$ and instellations $S_0$ (constant in time). Fig. \ref{fig:bithresh} represents the equilibrium states of these simulations, plotting the mantle melt fraction versus $F_c$ and $S_0$. For a given amount of tidal heat flux -- potentially regulated by an overlying atmosphere, as in Section \ref{ssec:res_lov} -- a planet would evolve from the left side to the right side of the plot, on the understanding that its instellation decreases over time. Unless the tidal heating is particularly efficient (large $F_c$), a wide range of these toy planets undergo a `blue sky bifurcation' upon intersection with the red line. This is the point at which their melt fraction decreases from $>30\%$ on the left side to less than $1\%$ on the right side. Mapping these semi-analytic coordinates ($S_0$,$F_c$) to time and $F_\text{tide}$ requires the application of our \textsc{proteus} framework.
\par 
Given that our \textsc{proteus} simulations (Section \ref{ssec:res_lov}) reproduce the tide-radiative feedback exhibited by these semi-analytic models, it is possible that these late stage bifurcations may extend to physically representative simulations, and thus to reality. For observations of some exoplanets consistent with the absence of tidally heated interiors, it could be the case that this is a late-stage characteristic of their evolution, having spent several Gyr with a molten interior. Realistically simulating these planets across Gyr timescales, with sufficiently small time-steps to maintain radiation-tide-rheology equilibrium, is currently beyond the computational limitations of the \textsc{proteus} modelling framework. 

\section{Atmospheric escape}
\label{app:escape}
The energy limit mass loss rate for escape  $\dot{M}_{EL}$ \citep{Watson1981, Lehmer_2020} is given by 
\begin{equation}
\left(\frac{\sqrt{2} R_{\text{optical}}}{R_{\text{xuv}}} \right)^3 \dot{M}_{EL}(t) = \frac{3\eta F_{\text{xuv}}(t)}{\sqrt{2} G \rho}, 
\label{eq:EL}
\end{equation}
where $F_{\text{xuv}}(t)$ is the time-evolving XUV flux, $R_{\text{optical}}$ is the optical transit radius of the planet, ${R_{\text{xuv}}}$ is the XUV absorption radius that contributes the effective planetary cross section to incoming XUV flux, $\rho$ is the bulk density of the planet, $G$ is Newton's gravitational constant and $\eta$ is the efficiency factor. The integration of the RHS of Equation \ref{eq:EL} with the XUV flux as a function of time and all else held constant allows calculation of a cumulative volatile loss quantity, corresponding to XUV fluence, a variable in the cosmic shoreline \citep{zahnle_cosmic_2017}. Hydrogen dominated atmospheres enriched in effective molecular coolants will have reduced escape efficies, such as $\eta=10\%$ \citep{yoshida_escape_2024}. \citet{Lehmer_2020} calculate that for a hydrogen envelope a few percent of planet mass, the increase in absorption radius of XUV can increase the escape rate by orders of magnitude. On balance, the quantity expresses the potential history of volatile loss could be for a given planet, while being simply derivable from the flux. To explore detailed questions of loss and revival \citep[e.g.][]{kite_exoplanet_2020}, self-consistent coupling of the escape to the \textsc{proteus} framework for $R_{\text{xuv}}(t)$ would be required.

\section{Outgassed atmospheric compositions}
\label{app:outgas}

In this appendix, we present the outgassed volatile atmospheric compositions calculated in our main simulation results (Section \ref{ssec:res_lov}) for completeness. Our modelled atmospheres are treated with constant mixing ratios, which are calculated according to our equilibrium outgassing scheme (Section \ref{sec:methods}).

\par
\begin{figure}
    \centering
    \includegraphics[width=0.98\linewidth, keepaspectratio]{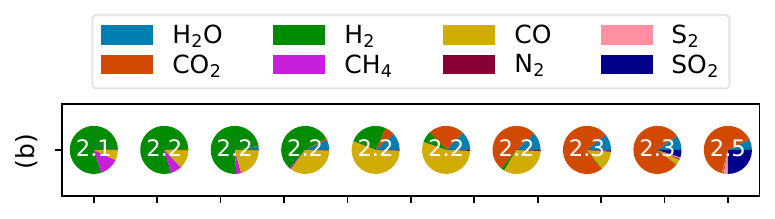}%
    \vspace*{-2mm}
    \includegraphics[width=0.98\linewidth, keepaspectratio]{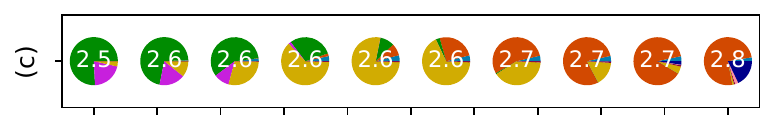}%
    \vspace*{-2mm}
    \includegraphics[width=0.98\linewidth, keepaspectratio]{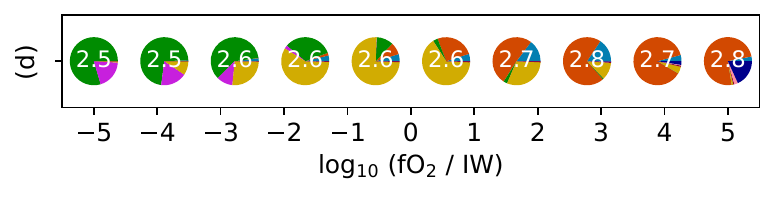}%
    \caption{Outgassed atmospheric composition at model termination for the three L 98-59 planets (panels) versus mantle oxygen fugacity (x-axis), for a core radius fraction $r_c=50\%$. These pie charts show volatile volume mixing ratios (wedge colours), while labelled white numbers show total surface pressure in log units: $\text{log}_{10} (P_\text{surf}/\text{bar})$.
    }
    \label{fig:outgas}
\end{figure}

 
Fig. \ref{fig:outgas} shows that outgassed atmospheres exhibit significant variations in composition across a range of mantle oxygen fugacities $f\ce{O2}$. Depending on $f\ce{O2}$, atmospheric composition shifts from \ce{H2}-dominated at reducing conditions, through \ce{CO}-dominated, to \ce{CO2}-dominated at more oxidising conditions. These regimes are consistent with behaviours described in the literature \citep{nicholls_redox_2024, nicholls_convective_2024,gaillard_diverse_2021,lichtenberg_review_2025,seidler_impact_2024}. We find that \ce{H2O} is never the dominant gas component because it remains favourably dissolved into the underlying magma ocean \citep{sossi_solubility_2023,boer_absence_2025}. In contrast, \citet{zahnle_moon_2015} assumed \ce{H2O}-dominated atmospheres. Total surface pressure (indicated by the white text) varies from $\sim125$ to $\sim\SI{630}{\bar}$. These optically-thick envelopes allow the atmosphere to regulate the planetary energy balance, and give rise to the radiation-tide-rheology feedback. 

\par 
We have nominally assumed (Section \ref{sec:methods}) an elemental inventory based on estimates of Earth's primitive mantle \citep{wang_elements_2018} in order to restrict the scope of our investigation. Earth is relatively volatile-poor \citep{korenaga_earth_2008, schaefer_review_2018}; larger atmospheres would provide more efficient blanketing of these planets' interiors, and make long-lived magma oceans increasingly feasible \citep{lichtenberg_vertically_2021, nicholls_redox_2024}. In reality, the volatile inventories of these planets will have been modulated by the complexities of planet formation, and then atmospheric escape over billions of years. These additional processes should be explicitly investigated in future work.

\bsp	
\label{lastpage}
\end{document}